\documentclass[letterpaper,twocolumn,10pt]{article}
\usepackage{usenix}

\pdfoutput=1

\usepackage{microtype}
\usepackage{lmodern}
\usepackage{multicol}
\usepackage[T1]{fontenc}
\usepackage[utf8]{inputenc}
\usepackage[english]{babel}
\usepackage{pdfpages}
\usepackage{stmaryrd}
\usepackage{graphicx}
\usepackage{tabularx}
\usepackage{xspace}
\usepackage[hyphens]{url}
\usepackage{multirow}
\usepackage{listings}
\usepackage{hyperref}
\usepackage{MnSymbol}
\usepackage{xtab}
\usepackage{caption}
\usepackage{enumitem}
\usepackage{balance}
\usepackage[normalem]{ulem}
\usepackage{float}
\restylefloat{table}
\usepackage{layouts}

\usepackage{caption}

\newcommand{\OURNAME}{\textsc{Android Security Framework}\xspace}
\newcommand{\OURSHORT}{\textsc{ASF}\xspace}
\newcommand{\KMODULE}{\textsc{Kernel Sub-Module}\xspace}
\newcommand{\MMODULE}{\textsc{Middleware Sub-Module}\xspace}

\definecolor{lightgray}{gray}{0.95}

\hypersetup{
    pdftitle={Android Security Framework: Enabling Generic and Extensible Access Control on Android},    
    pdfauthor={Michael Backes, Sven Bugiel, Sebastian Gerling, Philipp von Styp-Rekowsky},     
}

\date{}

\title{\OURNAME:\\Enabling Generic and Extensible Access Control on Android}

 \author{
 {\rm Michael Backes, Sven Bugiel, Sebastian Gerling, Philipp von Styp-Rekowsky}\\
 \{backes,bugiel,sgerling,styp-rekowsky\}@cs.uni-saarland.de\\
 Saarland University/CISPA, Germany
 } 

\begin{document}

\includepdf{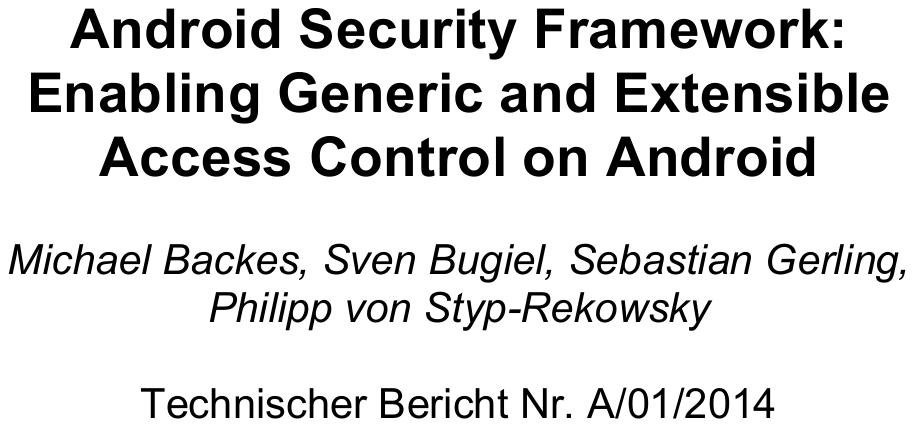}

\maketitle

\subsection*{Abstract}
We introduce the \emph{Android Security Framework~(ASF)}, a generic, extensible security framework for Android that enables the development and integration of a wide spectrum of security models in form of code-based security modules. The design of \OURSHORT reflects lessons learned from the literature on established security frameworks (such as Linux Security Modules or the BSD MAC Framework) and intertwines them with the particular requirements and challenges from the design of Android's software stack. \OURSHORT provides a novel security API that supports authors of Android security extensions in developing their modules. This overcomes the current unsatisfactory situation to provide security solutions as separate patches to the Android software stack or to embed them into Android's mainline codebase. As a result, \OURSHORT provides different practical benefits such as a higher degree of acceptance, adaptation, and maintenance of security solutions than previously possible on Android. We present a prototypical implementation of \OURSHORT and demonstrate its effectiveness and efficiency by modularizing different security models from related work, such as context-aware access control, inlined reference monitoring, and type enforcement.

\section{Introduction}
\label{sect:intro}
For several decades now, the need for operating system security mechanisms to provide strong security and privacy guarantees has been well understood~\cite{Linden:1976:OSS:356678.356682,1451869,Loscocco98theinevitability,Baker:1996:FBU:304851.304886}. Yet, recent attacks against smartphone end-user's privacy and security~\cite{Grace2012,Zhou:2013:NDSS,Zhou:2012:DAM:2310656.2310710,PorterFelt2011,PoWaMoHaCh_11:PermRedeleg,ChPoGrWa_11:AnalyzingInterApp,Felt:2012:APU:2335356.2335360} have shown that the fairly new smart device operating systems fail to provide these strong guarantees, for instance, with respect to access control or information flow control. To remedy this situation, security research has proposed a wide spectrum of security models and extensions for mobile operating systems, most of them for the popular open-source Android OS. These extensions range from context-related access control~\cite{CoNgCr_10:CRePE}, to developer-centric security policies~\cite{OnMcEnMc_09:Saint} and dynamic, fine-grained permissions~\cite{NaKhZh_10:Apex,backes13TACAS,Jeon2012,tissa11}, to domain isolation~\cite{trustdroid,BuDaDm_12:TowardsT}, and type enforcement~\cite{Smalley2013,TUD-CS-2013-0115}.

However, the lack of a comprehensive security API for the development and modularization of security extensions on Android has created the unsatisfactory situation that all of these novel and warranted security models are either provided as model-specific patches to the Android software stack, or they became an integrated component of the Android OS design~\cite{Smalley2013}. When considering the body of literature on established security frameworks, such as \textit{Linux Security Modules}~(LSM)~\cite{Wright:2002:LSM:647253.720287} or the \textit{BSD MAC Framework}~\cite{conf/usenix/WatsonMVF03}, their history has taught that the need to patch the OS or the hardwiring of a specific security model impairs both the practical and theoretical aspects of security solutions. First, there is in general no consensus on the \textit{``right''} security model, as demonstrated by the broad range of Android security extensions~\cite{CoNgCr_10:CRePE,OnMcEnMc_09:Saint,backes13TACAS,tissa11,trustdroid,BuDaDm_12:TowardsT,Smalley2013}. Thus, OS security mechanisms should not limit policy authors to one specific security model by embedding it into the OS design. Second, providing security solutions as \textit{``security-model-specific Android forks''} impedes their maintainability across different OS versions, because every update to the Android software stack has to be re-evaluated for and applied to each fork separately.

\paragraph{Contributions.}

In this paper, we propose the design and implementation of an \OURNAME~(\OURSHORT) that allows security experts to develop and deploy their security models in form of modules (or ``security apps''). This provides the means to easily extend the Android security mechanisms and avoids that policy authors have to choose ``the right Android security fork'' or that the OS vendor has to impose a specific security model. In the design of \OURSHORT we transfer the lessons learned and guiding principles from the literature on established OS security infrastructures to Android and intertwine them with new requirements for efficient security policies for multi-tiered software stacks of smart devices. This design solves a number of challenges in establishing a generic and extensible security framework for Android and we make the following concrete contributions:

\textbf{1.~Policy-agnostic, multi-tiered security infrastructure:} The security infrastructure must avoid committing to one particular security model and enable authors of security extensions to develop as well as deploy their solutions in form of code. This requires special consideration of Android's multi-tiered software stack and the dominant programming languages at each layer of this stack. For \OURSHORT we solve this by integrating security-model-agnostic enforcement hooks into the Android kernel, middleware and application layer and exposing these hooks through a novel security API to module authors.

\textbf{2.~Enabling edit automata policies:} Various Android security solutions realize edit automata policies that not only truncate but also modify control flows.
In \OURSHORT, the application layer and middleware hooks are specifically designed to allow module authors to leverage the rich semantics of Android's application framework and to implement their security policies as edit automata. This required a re-thinking of the ``classical'' object manager design from the literature by shifting the edit automata logic from the infrastructure into the security modules.

\textbf{3.~Instantiation of existing security models:} We demonstrate the efficiency and effectiveness of our \OURSHORT by instantiating different security models from related work on type enforcement~\cite{TUD-CS-2013-0115,Smalley2013}, context-related access control~\cite{CoNgCr_10:CRePE}, Chinese Wall policies~\cite{BuDaDm_12:TowardsT}, and inlined access control~\cite{backes13TACAS} as modules.

\textbf{4.~Maintenance benefits for security extensions:} Our ported security modules show how \OURSHORT simplifies maintainability of security extensions across different OS versions by shifting the bulk of effort to the security framework maintainer. This is similar to the maintenance of the application framework for regular apps. Hence, a comparable benefit to regular apps~\cite{jisc} in adaption and stability across OS versions can be expected of security modules.

\textbf{5.~Research and development benefits:} We postulate that developing security solutions against a well documented security API also greatly contributes to \textit{a)}~a better understanding and analysis of new security models that form a self-contained unit instead of being integrated to various components of the Android software stack, \textit{b)}~a better reproducibility and dissemination of new solutions since modules can be easily shared and instantiated, and \textit{c)}~a more convient application of security knowledge to the Android software stack without the requirement to be familiar with the deep technical internals of Android.

\paragraph{Outline.}

The remainder of this paper is structured as follows. In Section~\ref{sec:background} we provide necessary technical background information on Android's design and security philosophy. We survey closest related work in Section~\ref{sec:relatedwork} and derive from this design principles for a generic security framework on Android in Section~\ref{sec:requirements}. We present the design and implementation of \OURSHORT in Section~\ref{sec:architecture} and show the instantiations of different security models from related work in Section~\ref{sec:usecases}. In Section~\ref{sec:eval} we evaluate our framework in terms of performance impact and discuss limitations of our approach. We conclude in Section~\ref{conclusion}.

\section{Background on Android}
\label{sec:background}
In this section we provide necessary technical background information on Android.

\subsection{Primer on Android}
\label{sec:primerandroid}

Android is an open-source software stack for embedded devices. The lowest level of this stack consists of a Linux kernel responsible for elementary services such as memory management, device drivers, and an Android-specific lightweight inter-process communication called Binder. On top of the kernel lies the extensive Android middleware, consisting of native libraries (e.g., SSL) and the application framework. System services in the middleware implement the bulk of Android's application API (e.g., the location service) and pre-installed system apps at the application layer, like Contacts, complement this API.

Although application layer and middleware apps and services are commonly written as Java code, they are compiled to \textit{dex} bytecode and run inside the \textit{Dalvik Virtual Machine}~(DVM). In addition to dex bytecode, apps and services can use native code libraries (i.e., C/C++) for low-level interactions with the underlying Linux system. Native code can be seamlessly integrated into dex bytecode by means of the \textit{Java Native Interface}.

Android apps are generally composed of different components. The four basic app components are \textit{Activities} (GUI for user interaction), \textit{BroadcastReceivers} (mailbox for broadcast \textit{Intent} messages), \textit{ContentProviders} (SQL-like data management), and \textit{Services} (long running operations without user interaction). All components can be interconnected remotely across application boundaries by using different abstractions of Android's Binder IPC mechanism, such as \textit{Intent} messages.

\subsection{Android's Security Philosophy}
\label{sec:androidsec}

Android's security philosophy dictates that all apps are sandboxed by executing them in separate processes with distinct user IDs~(\textit{UID}) and assigning them private data directories on the filesystem.

\begin{figure}[t]
  \centering
  \includegraphics[width=.85\linewidth]{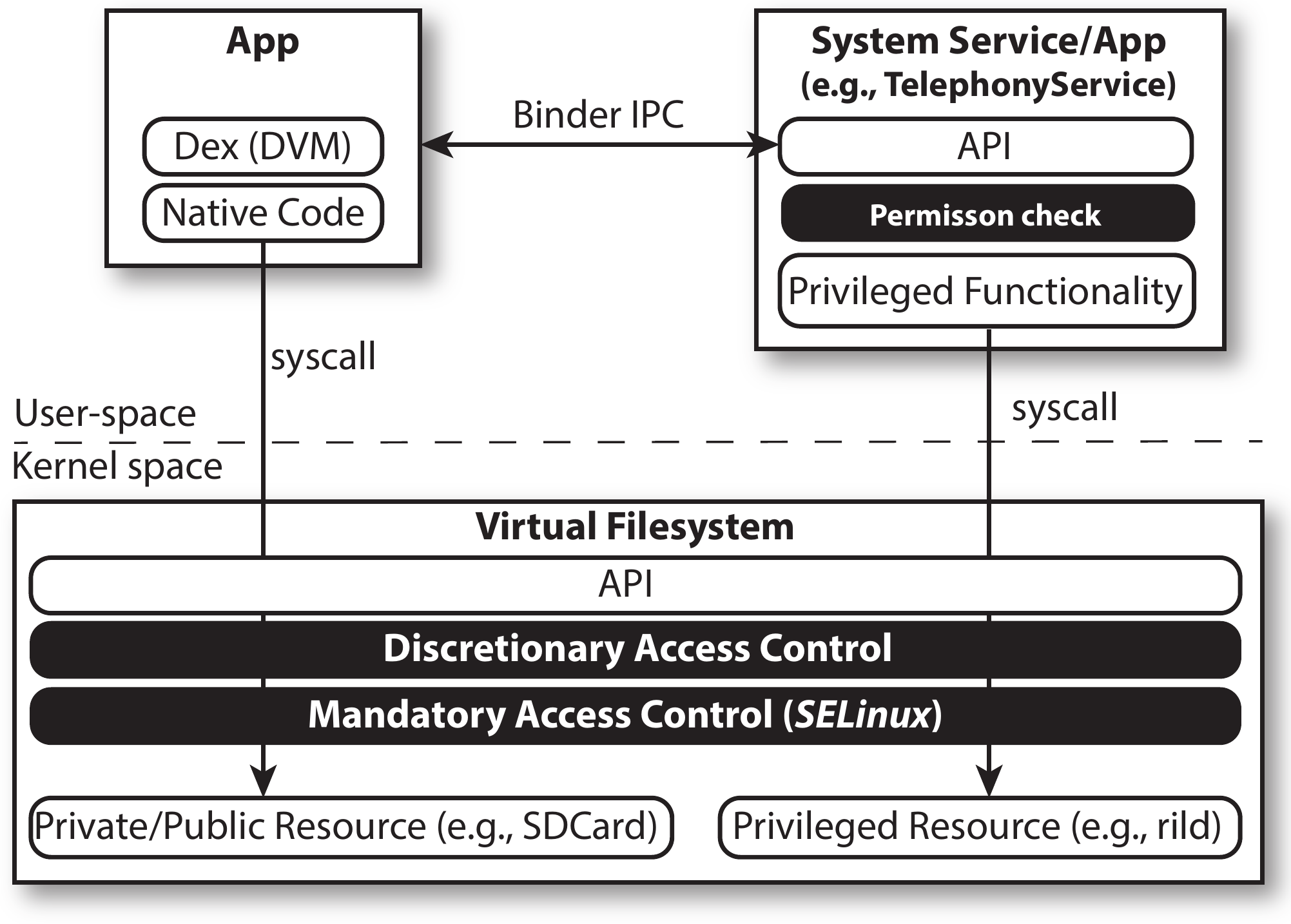}
  \caption{Android's security architecture.}
  \label{fig:background:secarch}
\end{figure}

To achieve privilege separation between apps, Android introduces \textit{Permissions}, i.e., privileges that an app is granted by the user at install-time. Without any permissions an app is not able to access security and privacy sensitive resources. Permissions are assigned to the app's UID and enforced at two different points in the system architecture, as depicted in Figure~\ref{fig:background:secarch}: First, every application sandbox can directly interact with the underlying kernel through system calls, for instance, to edit files or open a network socket. These resources are either of private nature (i.e., are within the app's private directory) or public resources (e.g., SDCard). Access control in the filesystem ensures that the apps' processes have the necessary rights (i.e., Permissions) to issues particular syscalls, e.g., to open a file on the SDCard. The filesystem access control consists of the traditional Linux Discretionary Access Control, which is complemented (since Android v4.3) by SELinux based Mandatory Access Control~(MAC).

Second, apps can interact through the Android API in a strictly controlled manner with highly privileged resources. To ensure system security and stability, apps are prohibited to access these highly privileged resources directly. Instead, those resources are wrapped by system services and apps that implement the API. For instance, the \textit{TelephonyService} communicates on behalf of apps with the radio interface layer daemon~(\textit{rild}) to initiate calls or send text messages. Whether an app is sufficiently privileged to successfully call the API is determined by a Permission check within the system services/apps. For this check, the Binder mechanism provides to the callee~(system service/app) the UID of the caller~(app).

\section{Related Work}
\label{sec:relatedwork}
We first provide a synopsis of the development of extensible kernel security frameworks and discuss afterwards the current status of security extensions and frameworks for the Android software stack.

\subsection{Extensible Kernel Access Control}
\label{sec:related:extkernel}

The importance of the operating system in providing system security has been very well studied in the last decades~\cite{1451869,Linden:1976:OSS:356678.356682,Baker:1996:FBU:304851.304886,Loscocco98theinevitability} and different approaches to extending operating systems with access control and security policies have been explored. These include system-call interposition~\cite{Fraser:2001:LMY:647054.715751,Provos:2003:IHS:1251353.1251371}, software wrappers~\cite{interposition1}, and extensible access control frameworks like \textit{Domain and Type Enforcement}~(DTE)~\cite{Badger:1995:PDT:882491.884237}, \textit{Generalized Framework for Access Control}~(GFAC)~\cite{gfac}, and \textit{Flask}~\cite{Flask99}. To realize these solutions, DTE has been provided as a patch to the UNIX system~\cite{Badger:1995:DTE:1267591.1267603}, while GFAC and Flask have been implemented as patches to the Linux kernel by the RSBAC~\cite{ott01} and SELinux~\cite{SELinux01} projects. However, this led to an intricate situation: On the one hand, maintaining these solutions as patches incurred high maintenance costs for adapting the patches to kernel changes. On the other hand, none of these solutions was included in the vanilla kernel because this would constrain security policy authors to one specific security model. This constrain would be unsatisfying since there exists in general no consensus on the ``right'' security model. To remedy this situation, extensible security frameworks have been proposed~\cite{Wright:2002:LSM:647253.720287,conf/usenix/WatsonMVF03} that allow the extension of the system with trusted code modules that implement specific security models. Module authors are supported with an API that exposes kernel abstractions as well as operations and facilitates the implementation of the desired security architecture and model. The results of this research have been integrated into the mainline kernels as the \textit{Linux Security Modules} framework~(LSM)~\cite{Wright:2002:LSM:647253.720287} and the \textit{BSD MAC Framework}~\cite{conf/usenix/WatsonMVF03}. Additionally, different access control models, such as SELinux type enforcement~\cite{smalley:01} or TOMOYO path-based access control~\cite{tomoyo}, have been ported as modules on top of the LSM and the BSD MAC framework.

\subsection{Android Security}
\label{sec:related:androidsec}

Security research and reported attacks have shown that the stock Android security mechanisms are insufficient to protect the end-user's privacy and the system security. At the same time, research has also proposed different security extensions for Android to mitigate these attacks and to improve protection of the end-user's privacy. The spectrum of the proposed solutions covers a wide range of security models and architectures. To name a few: \textit{CRePE}~\cite{CoNgCr_10:CRePE} provides a context-related access control, where the context can be, e.g., the device's location. \textit{Saint}~\cite{OnMcEnMc_09:Saint} enables developer-centric policies that allow app developers to ship their apps with rules that regulate the app's interactions with other apps and can thus protect the app from misuse and attacks. Different approaches to more dynamic and fine-grained permissions have been proposed based on system-centric enforcement (\textit{Apex}~\cite{NaKhZh_10:Apex} and \textit{TISSA}~\cite{tissa11}) or inlined reference monitors (\textit{Dr.~Android and Mr.~Hide}~\cite{Jeon2012}, \textit{Aurasium}~\cite{Xu2012a}, and \textit{AppGuard}~\cite{backes13TACAS}). \textit{XManDroid}~\cite{BuDaDm_12:TowardsT} enforces Chinese Wall policies to prevent confused deputy and collusion attacks. \textit{TrustDroid}~\cite{trustdroid} and \textit{MOSES}~\cite{Russello:2012:MSO:2295136.2295140} isolate different domains such as ``Work'' and ``Private'' from each other. \textit{SE~Android}~\cite{Smalley2013} and \textit{FlaskDroid}~\cite{TUD-CS-2013-0115} bring type enforcement to Android, where SE~Android focuses on the kernel layer and has been partially included into the mainline Android source code, and FlaskDroid extends type enforcement to Android's middleware layer on top of SE~Android.

\section{Motivation and Requirements Analysis}
\label{sec:requirements}
The current development of Android security extensions has strong parallels to the initial development of the above mentioned Linux and BSD security extensions, since current Android security extensions are provided as patches to the software stack or, in the case of SE~Android~\cite{Smalley2013}, are embedded into the Android source tree. For the same, above mentioned reasons as for the early Linux and BSD security extensions, this impedes the applicability and adaption of Android security extensions and additionally precludes many of the benefits that a modular composition could offer in terms of maintenance: Embedding SE~Android's security model into Android's source tree limits policy authors to the expressiveness and boundaries of type enforcement, whereas provisioning security models and architectures as patches to Android's software stack forces policy authors to chose a solution-specific Android fork. This requires for every version update to the Android OS a re-evaluation and port of each separate fork. Moreover, security solutions cannot be easily compared with each other, because their infrastructures are deeply embedded into the Android software stack.

In this paper, we develop in the spirit of the two de facto most established security frameworks, \textit{Linux Security Modules}~(LSM)~\cite{SELinux01} and the \textit{BSD MAC Framework}~\cite{conf/usenix/WatsonMVF03,UCAM-CL-TR-818}, a generic and extensible \OURNAME that allows the instantiation and deployment of different security models as loadable modules at Android's application layer, middleware, and kernel. The two most important guiding principles from LSM and the BSD MAC framework that govern the design of our \OURNAME are: \textit{1)}~provisioning of policies as code instead of data; and \textit{2)}~providing a policy-agnostic OS security infrastructure. In the remainder of this section, we analyze the requirements and challenges for their transfer to the Android software stack.

\paragraph{Policy as code and not data.}

The first guiding principle is that policies should be supported as code instead of data (such as rules written in one predetermined policy language). Providing an extensible security framework that supports loading of policy logic as code avoids committing to one particular security model or architecture. For Android, this removes the need to chose a particular extension-specific Android fork or to be limited to one specific security model in the mainline Android software stack. Additionally, developing modules against an OS security API provides the benefits of modularization for developing and maintaining security extensions. This includes, foremost, a higher functional cohesion of security modules and lower coupling with the Android software stack and, hence, can significantly reduce the maintenance overhead of modules, especially in case of OS changes. Moreover, it allows a better dissemination, comparison, and analysis of self-contained security modules.

Transferring this principle to an extensible security framework for Android poses the additional requirement to consider the semantics and dominant programming languages of the different layers of Android's software stack. LSM and the BSD MAC Framework, for instance, as part of the Linux and BSD kernels, support modules written in C and operate on kernel data structures (e.g., filesystem inodes). While this applies to the Android Linux kernel as well, an Android security framework should additionally support modules written for Android's semantically-rich middleware and application layers. That means modules written in Java and operating on application framework classes (e.g., Intents or app components).

\paragraph{Policy-agnostic security infrastructure.}

The second principle is that the security framework and its API should be policy-agnostic. This means that the different layers of the software stack are aware of the security infrastructure, but policy-specific intrusions into these layers are avoided and policy-specific data structures and logic are confined to security modules.

A particular additional requirement for a security framework on Android are enforcement hooks in the middleware and application layer that support \textit{edit automata}~\cite{LiBaWa_05:EditAutomata} policies, as promoted by different solutions~\cite{tissa11,HoHaJuScWe_11:RetrofittingAndroid,trustdroid,Jeon2012,backes13TACAS}. Edit automata, in contrast to truncation automata, can not only abort control flows but also divert or manipulate them and, thus, give policy authors a higher degree of freedom in implementing thir enforcement strategies. For instance, when querying a ContentProvider component, the policy could simply deny access by throwing a Java SecurityException (truncation), but also modify the return value to return filtered, empty, or fake data~(edit). To technically enable security modules to implement edit automata, our design requires a re-thinking of the ``classical'' object manager vs.~policy server design that is used, e.g., in LSM. Object managers (i.e., enforcement points) are responsible for assigning security labels to the objects that they manage and for both requesting and enforcing access control decisions from the policy server (i.e., policy decision point). Because this design embeds the enforcement logic into the system independently from the security model, it is unfit for realizing edit automata. Thus, our design requires hooks that generically support different enforcement strategies and shift the enforcement and object labelling logic from the object managers to the security modules.

Moreover, the security framework should provide a policy-agnostic infrastructure for common operations such as event notifications or module life-cycle management---thus reducing the overhead for policy authors to implement common functionality.

\section{Android Security Framework}
\label{sec:architecture}
\begin{figure*}[t]
  \centering
  \includegraphics[width=.65\linewidth]{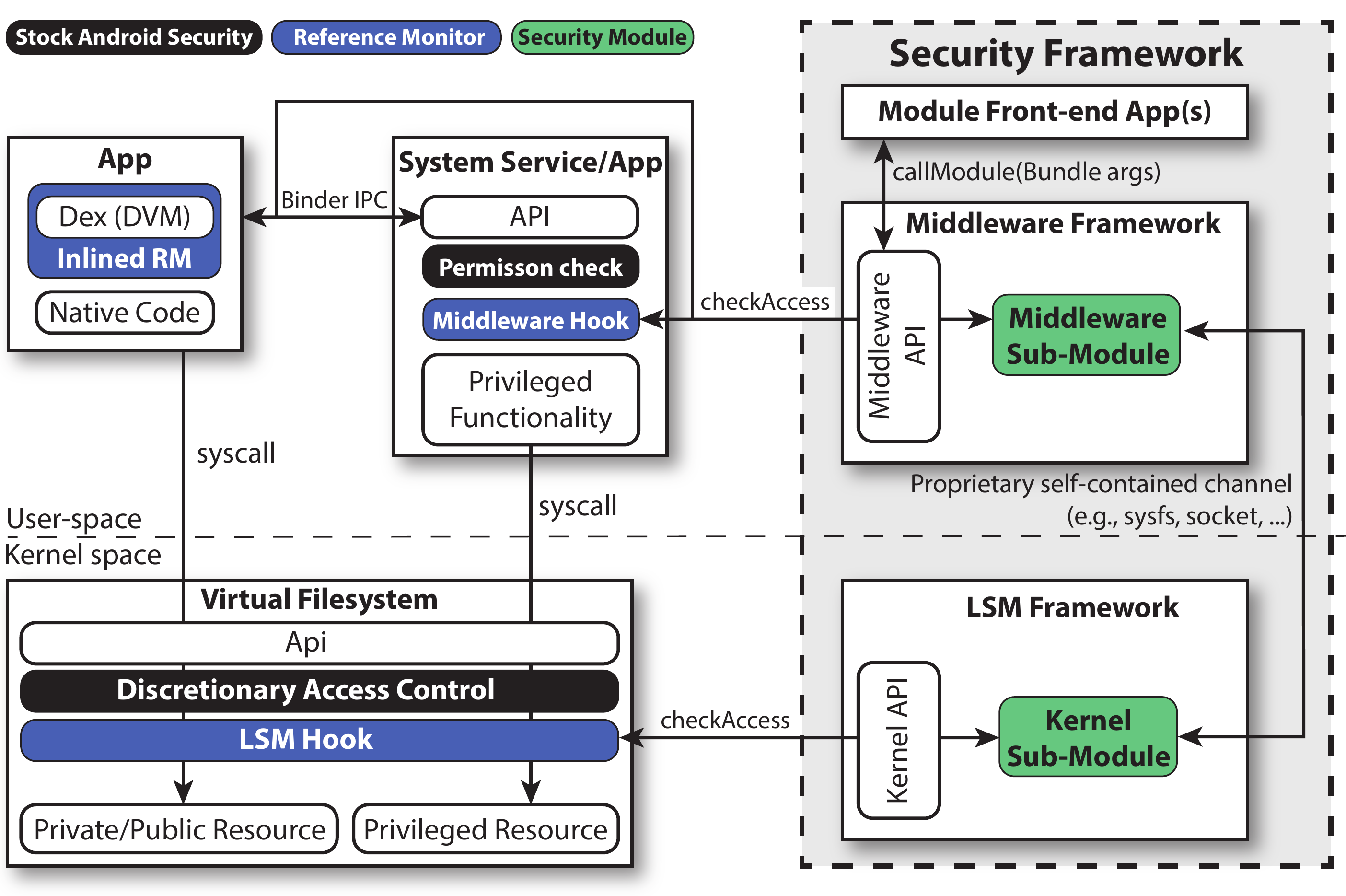}
  \caption{Android Security Framework architecture.}
  \label{fig:arch}
\end{figure*}

In the following we present \OURSHORT. We provide further technical details in the appendices of this submission.

\subsection{Framework Overview}
\label{sec:arch:overview}

The basic idea behind our \OURNAME is to extend Android with a new security API that incorporates the design principles explained in Section~\ref{sec:requirements}. This API allows to easily author, integrate, and enforce generic security policies. Figure~\ref{fig:arch} provides an overview of our \OURSHORT and we explain its building blocks in the following.

\subsubsection{Reference Monitors}
In our design we differentiate between policy enforcing code and policy decision making code. For enforcement we use reference monitors~\cite{Lampson:1974:PRO:775265.775268,orangebook} at all layers of the Android software stack, i.e., at the application layer, the middleware layer, and the kernel layer. Each reference monitor protects one specific privileged resource and is placed such, that it is always invoked by the control flow between the Android API and access to the resource. The benefit of this multi-tiered enforcement is that each reference monitor can operate with the semantics of its respective layer. In conjunction all monitors enable exceedingly powerful and semantically rich security policies.

\subsubsection{Security Modules}
Security extensions are deployed in the form of code modules and loaded into the security frameworks at the middleware and kernel level. Each module implements a policy engine that manages its own security policies and acts as policy decision making point. Security modules are integrated into the security frameworks through a security API that exposes objects and operations of the different software stack layers. As part of this API, each module implements an interface for enforcement functions (i.e., functions that make a policy decision for a particular reference monitor in the system), management functions (e.g., module life-cycle events), and other interfaces that will be explained in Section~\ref{sec:arch:framework}.

To provide a clear separation between policy decision logic using kernel level semantics and logic using middleware/application layer semantics, each module consists of two sub-modules: a \KMODULE leveraging the already existing Linux Security Module~(LSM) infrastructure of the Linux kernel and a \MMODULE, for which we designed and implemented a novel security infrastructure at the application and middleware layers.

\subsubsection{Front-end Apps}
To enable user configurable policies or graphical event notifications, modules might want to include user interfaces. To this end, the module developers (or external parties being aware of the modules) can deploy standard Android apps that act as front-end apps and communicate through the framework API directly with the module. In our framework API, we enable this communication through a \textit{Bundle} based communication protocol. A Bundle is a key-value store that supports heterogenous value types (e.g., Integer and String) and that can be transmitted via Binder IPC between the app process and the module. It is the responsibility of the module to verify that the caller has the necessary privileges to issue commands.

\subsection{Framework Infrastructure}
\label{sec:arch:framework}

We present now in a bottom-up approach details about the \OURSHORT infrastructure that has been prototypcially implemented for Android v4.3 and currently comprises 4606 lines of code.

\subsubsection{Kernel Space}

At the kernel level we employ the existing Linux Security Module~(LSM)~\cite{Wright:2002:LSM:647253.720287} framework of the Linux kernel. LSM implements an infrastructure for mandatory access control at kernel level and provides a number of enforcement hooks within kernel components such as the process management, the network stack, or the virtual filesystem. The \KMODULE is implemented as a standard Linux Security Module that registers through the LSM API for the LSM hooks in the system. This enables the implementation of a variety of different security models that operate with kernel level semantics. Modules include, for instance, SELinux type enforcement~\cite{smalley:01}, which uses security labels attached to filesystem inodes and process data structures, or TOMOYO~\cite{tomoyo} for path-based access control, or custom modules~\cite{nadkarni:ccs13}. Kernel-level policies form truncation automata that terminate illegal control flows, e.g., on access to files.

Since there might be operational inter-dependencies between the \KMODULE and user-space processes (e.g., re-labelling the security context of files or propagating kernel access control decisions to the \MMODULE), the kernel module can implement proprietary channels for communication between kernel- and user-space (e.g., sockets or sysfs entries).

\subsubsection{Middleware Layer}

At the middleware layer we extended the system services and apps that implement the Android API with hooks that enforce access control decisions made by the \MMODULE. The middleware security framework is executed as a new Android system service and mediates between our hooks and the \MMODULE. In contrast to previous solutions for (generic) access control on Android, our hooks are policy-agnostic and not tailored to one specific security model. Each hook takes as arguments all relevant, ambient information of the current control flow that led to the hook's invocation. For instance, Listing~\ref{listing:hookexamples} presents two exemplary hooks in our system: one for the Intent broadcasting subsystem of the \textit{ActivityManagerService} (line 1) and one for the \textit{LocationManagerService} that implements the location API of Android (line 2). Both provide to the \MMODULE information about the current caller to the Android API, i.e., \textsc{App} in Figure~\ref{fig:arch} (parameters \texttt{callingUid} and \texttt{callingPid}). However, all other parameters are specific to the hooks' contexts, e.g., the hook in line 1 provides information about the Intent being broadcasted (parameter \texttt{intent}) and the app component that should receive this Intent (parameters \texttt{targetComp} through \texttt{targetPid}). Thus, the hooks support policies that use the rich middleware-specific semantics.

In general, all hooks support truncation automata as policies by either allowing the module to throw exceptions that terminate the control flow and that are returned to the caller of the Android API, or by explicitly requiring a boolean return value that indicates whether the hook truncates the control flow or not (line 1 in Listing~\ref{listing:hookexamples}). A subset of the hooks additionally supports edit automata policies, that is the module can modify or replace return values of the Android API function or modify/replace arguments that divert or affect the further control flow after the hook. For instance, the \textit{LocationManagerService} hook in Listing~\ref{listing:hookexamples} (line 2) allows the module to edit or replace the Location object that is returned to the app that requested the current device location.

\begin{lstlisting}[caption={Exemplary enforcement functions}, label={listing:hookexamples}, float, basicstyle=\footnotesize, language=Java]
public boolean security_broadcast_deliverToRegisteredReceiver (Intent intent, ComponentName targetComp, String requiredPermission, int targetUid, int targetPid, String callerPackage, ApplicationInfo callerApp, int callingUid, int callingPid);
public Location security_location_getLastLocation (Location currentLocation, LocationRequest request, int callingUid, int calingPid);
\end{lstlisting}

We describe the API of the middleware framework as well as the detailed structure of modules in the next sections. Appendix~\ref{sec:appendix:coverage} provides an overview of the coverage of our current enforcement hooks.

\subsubsection{Application Layer}

At the application layer, our \OURNAME provides a mechanism to inject access control hooks into apps themselves. This access control technique is based on the concept of \emph{inlined reference monitors} (IRM) pioneered by Erlingsson and Schneider~\cite{ErSc_00:IRMEnforcement}. The basic idea is to rewrite an untrusted app such that the reference monitor is directly embedded into the app itself, yielding a ``self-monitoring'' app. The main advantage of policy enforcement in the caller's process context is that the hook and transitively the security module has full access to the internal state of the app and can thus provide rich contextual information about the caller. For instance, by inspecting the call stack, the IRM is able to distinguish between calls from different app components, e.g., advertisement libraries, thereby allowing more expressive and fine-grained access control policies. \OURSHORT provides an instrumentation API that enables security modules to dynamically hook any Java function within an app's DVM. Hooked functions divert the control flow of the program to the reference monitor, which thereby not only gains access to all function arguments but can also modify or replace the function's return value. Thus, the IRM is also able to enforce edit automata security policies. Furthermore, in contrast to the hooks placed in the Android middleware, application layer hooks are dynamic: Hooks are injected by directly modifying the target app's DVM memory when a new app process is started. This design enables security modules to dynamically create and remove hooks at runtime as well as to inject app-specific hooks.

\subsection{Middleware Framework API}
\label{sec:arch:api}

We elaborate now in more detail on our framework API and the interaction between modules and the security infrastructure. Since we use the LSM framework as is, we focus here on our newly introduced middleware security framework and refer to the kernel documentation~\cite{lsmdocu} for details on the LSM API.

The middleware framework API of our current implementation contains 168 functions. A full listing of our current API is provided in Appendix~\ref{sec:appendix:api}. This API can be broken down into the following categories:

\paragraph{Enforcement functions.} These module functions form the bulk of the API (136 methods) and are called by the framework whenever the enforcement hooks in system apps and services are triggered. Each hook has a corresponding function in the module API that implements the policy decision logic for this hook. Enforcement functions have the same method signatures as their hooks (cf.~Listing~\ref{listing:hookexamples}), i.e., all parameters provided by a hook are passed to its enforcement function. Passing arguments by reference or expecting objects as return values allows these functions to implement edit automata logic.

\paragraph{Kernel Sub-Module Interface.}

To avoid policy-specific interfaces for the communication between middleware/application layer apps and the \KMODULE, we introduce a generic kernel module API as part of the middleware framework API. It allows apps and services a controlled access to Linux security modules (cf.~Listing~\ref{listing:kmacapi} in Appendix~\ref{sec:appendix:api}). Each security module can implement this interface and internally translate the API calls to calls on the proprietary channel between the user-space and the Linux security module. Two particular challenges for establishing this interface were the self-contained security checks of the kernel module and the requirement that this interface is already available during system boot. To guarantee security, the kernel module is required to perform policy checks to verify that a user-space process is sufficiently privileged to issue commands to it. Additionally, the kernel module is called before the middleware framework can load any \MMODULE, e.g., it can be called by Zygote when spawning new app processes. To solve these challenges, our design avoids an additional layer of indirection (i.e., IPC) for communication with the kernel module and loads the interface implementations via the Java reflection API statically into the application framework when it is bootstrapped. This ensures that the calling processes communicate directly with the kernel module through our generic API and that the kernel module can be called independently of middleware services.

\paragraph{Life-cycle management.}

Every module must implement functions for life-cycle management, such as initialization or shutdown. This enables the framework to inform the module when the system has reached a state during the boot cycle from which on the module will be called or when the system shuts down. Modules should use these functions, e.g., to initiate their policy engines or to save internal states to persistent storage before the device turns off.

\paragraph{Event notifications.}

Event notification interfaces are used to propagate important system events to the module. For instance, modules should be immediately informed when an app was successfully installed, replaced, or removed. Although this information is usually propagated via a broadcast Intents, the time gap between package change and broadcast delivery might cause inconsistencies in module states. Hence these events must be delivered synchronously.

\paragraph{Framework Callbacks.}

The framework provides modules a callback interface for communicating in a more direct manner with system services, such as the \textit{PackageManagerService}, and avoids the need to go through the Android API. This is desirable for policy authors that want to leverage the middleware internal information for a more efficient access control enforcement. Our current callback interface, for instance, includes functions that allow modules to efficiently resolve PIDs to application package names.

\paragraph{Proprietary protocols.}

We introduced in our framework API a \textit{callModule()} function that allows modules to implement proprietary communication protocols with other apps that are aware of this specific module, e.g., the front-end apps (cf.~Section~\ref{sec:arch:overview}). When using \textit{callModule()}, these protocols are based on \emph{Bundles} and enable a protocol similar to the Parcel-based Binder IPC: apps serialize function arguments to a Bundle and add an identifier for the function the receiver should execute with the deserialized arguments.

\paragraph{IRM Instrumentation.}

The framework provides an instrumentation API that enables security modules to hook any Java function within selected app processes. Function calls can be redirected to an inlined reference monitor that enforces policy decisions made by the module. Appendix~\ref{sec:appendix:irm} explains an example that uses the instrumentation API.

Hooks injected via the instrumentation API are local to the app process that the API is called from. Therefore, all calls to the instrumentation API need to be performed from within a target application's process. We solve this by placing an instrumentation hook in the \textit{ActivityManagerService} that is triggered when a new app process is about to be launched. A module that implements this hook has to return the name of a module's Java class (the inlined reference monitor) that will be executed within the app's process before control flow is passed to the app itself. This is accomplished by modifying the arguments passed to Zygote: Instrumented apps are started via a special wrapper class that loads and executes the instrumentation code before running the app.

\subsection{Middleware Security Modules}
\label{sec:arch:modules}

\begin{figure}[t]
  \centering
  \includegraphics[width=.6\linewidth]{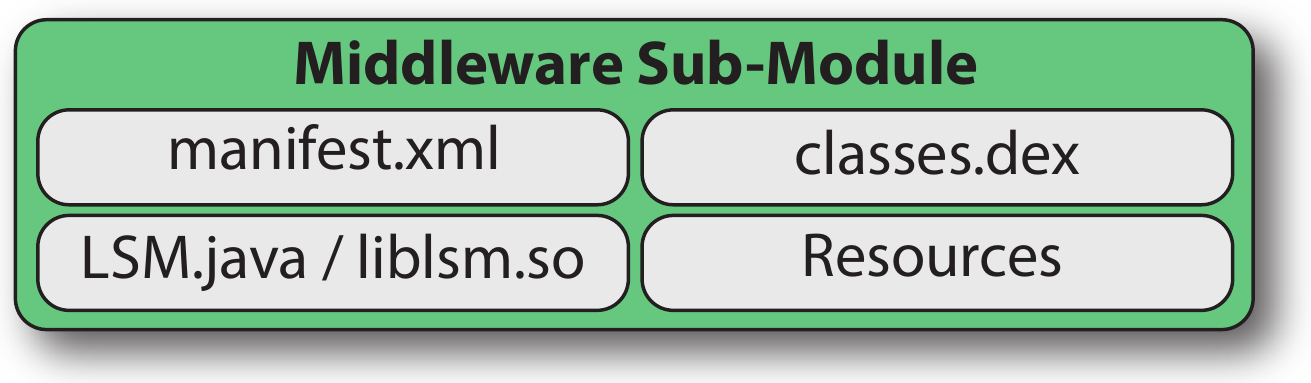}
  \caption{Middleware security module structure.}
  \label{fig:arch:modulearch}
\end{figure}

We elaborate in more detail on the structure of security modules. Again, we use Linux security modules as is~\cite{lsmdocu} and, thus, focus here on the \MMODULE. A middleware security module is, simply speaking, an app that is created with an Android SDK that includes our new security API. It is deployed to a protected location on the file system, from where it is loaded during boot. The module package is a \emph{Jar} file that contains all the module's program code, resources, and manifest file (cf.~Figure~\ref{fig:arch:modulearch}):

\paragraph{Module Manifest.} The manifest (formatted in XML) declares properties such as the module author or code version, and, more importantly, the name of the main Java class that forms the entry point for the module.

\paragraph{Classes.dex.} The \textit{classes.dex} file contains, as in regular Android apps, the Java code compiled to \textit{Dalvik executable bytecode}~(DEX). It contains all Java classes that implement the security module's logic. During the load process of the \MMODULE, the middleware framework uses the Java reflection API to load the module's main class (as specified in the manifest) from \textit{classes.dex}. To ensure that the reflection works error-free, the main class must implement the API as described in Section~\ref{sec:arch:api} (and listed in Appendix~\ref{sec:appendix:api}). Since the API defines currently more than a hundred methods, but a security module very likely requires only a subset of those, our SDK provides an abstract class that implements the API. That abstract class can be sub-classed by the module's main class, which then only needs to override the required functions. The abstract class returns for each non-overridden enforcement function an allow decision.

\paragraph{LSM interface.} The proprietary interface between the user-space processes and the Linux security module in the kernel is implemented through a native library \textit{liblsm.so} and a corresponding Java class \textit{LSM.java}, which exposes the native library via the Java Native Interface. \textit{LSM.java} has to implement the generic interface for the communication with the kernel that was explained in the previous section. The generic kernel module interface of \OURSHORT (called \textit{KMAC.java}) loads \textit{LSM.java} through the Java reflection API into Android's application framework. This allows apps and services to communicate via \textit{KMAC} (and reflectively through \textit{LSM.java}) with the kernel module and avoids a policy-specific interface. We exemplify this mechanism in Appendix~\ref{sec:appendix:kmacexample} by integrating SELinux into Zygote.

\paragraph{Resources.} Each module can ship with proprietary resources, such as initial configuration files or required binaries (e.g., \textit{ccstools} for TOMOYO~\cite{tomoyo}). During module instantiation, the framework informs the module about the filesystem location of its Jar file, enabling the module to extract these resources on-demand from its file.

\subsection{Support for Stackable and Dynamical Loadable Modules}
\label{sec:eval:loadable}

Finally, two desirable properties for implementing an extensible security framework such as our \OURSHORT are dynamically loadable policies and policy composition (i.e., stacking modules). In the following we explain why we chose to \textit{permit} these features by design, but \textit{not} consider them a requirement for our solution.

\paragraph{Dynamically Loadable Modules.} Being able to dynamically load and unload modules is desirable, for instance, to speed up the development and testing cycles of modules and, in fact, we used this feature during the development of our example use-cases (cf.~Section~\ref{sec:usecases}). However, the arguments to support dynamically loadable modules beyond development are disputed: First, dynamic loading is not always technically possible. A small set of static policy models, such as type enforcement~\cite{Smalley2013,TUD-CS-2013-0115}, require that all subjects and objects are labeled with a security context. Supporting such extensive labeling operations at runtime is an intricate problem. Second, there exist security considerations. The loading and unloading of modules must be strictly controlled to ensure that only integrity protected, trusted modules are loaded. Otherwise, given the privileges of modules, this would open the way to powerful malware modules. In our design we agree with the conclusions of the various Linux security module authors~\cite{loadablelsm} and consider the drawbacks of dynamically loadable modules to outweigh their benefits. Therefore, we load the module once during the system boot and \emph{permit} users of our framework to additionally activate dynamic unloading and loading of modules. But we currently do not consider this feature a requirement for our solution. However, community and market forces (e.g., vendors of security solutions) have to determine whether there is a need for supporting dynamically loadable modules in the future (e.g., establishing a ``security app market'') and, hence, whether we have to revise our requirements analysis.

\paragraph{Stackable Modules.} Composing the overall policy from multiple, simultaneously loaded and independent policies is a desirable feature, since usually no ``one-size-fits-all'' policy exists. Android, for instance, implements currently a quadruple-policy approach consisting of Permissions, SE~Android type enforcement, AppOps, and Linux capabilities---each being responsible for a different aspect of the overall access control strategy. Multiple policies will naturally conflict and thus require the security framework to support different policy composition and reconciliation strategies (e.g., consensus or priority based)~\cite{Rao:2009:DMA:1542207.1542217,1004363}. However, supporting fully generic policy composition is quite a challenge and has been shown to be intractable~\cite{674833}. Thus, despite its benefits, we decided in our design to follow the lessons learned by the LSM developers~\cite{Wright:2002:LSM:647253.720287} and to only \emph{permit} module developers to implement stackable modules, but we do not provide explicit interfaces for stacked modules in our framework infrastructure. The approach to stacking modules would be to provide a ``composition module'' that implements policy reconciliation and composition logic and which in turn can load other modules and multiplex API calls between them.

\section{Example Security Modules}
\label{sec:usecases}
\begin{figure*}[t]
\centering
\begin{minipage}[b]{.38\linewidth}
  \centering
  \includegraphics[width=\linewidth]{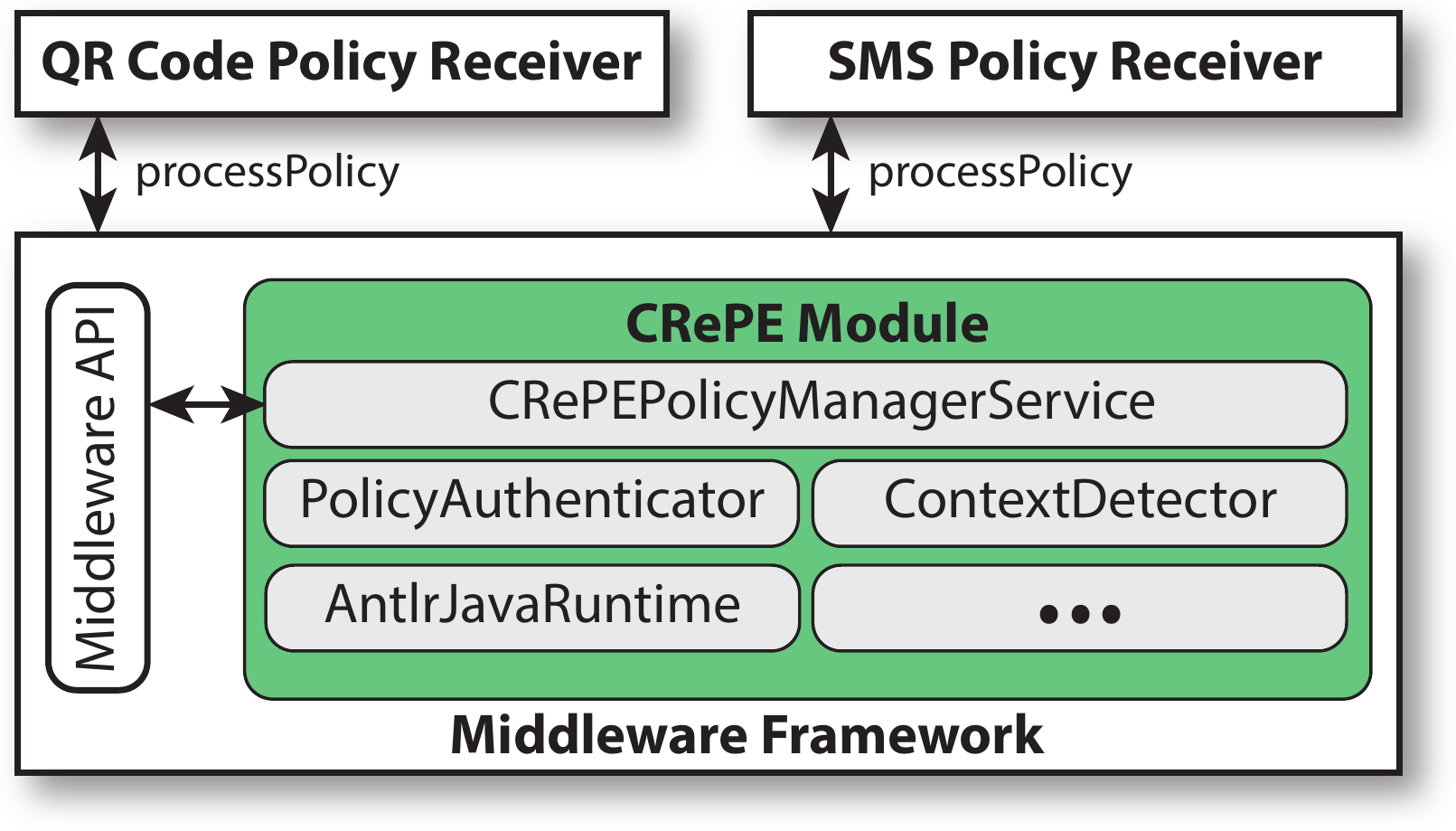}
  \caption{CRePE module}
  \label{fig:crepemodule}
\end{minipage}
\hfill
\begin{minipage}[b]{.3\linewidth}
\centering
  \includegraphics[width=.9\linewidth]{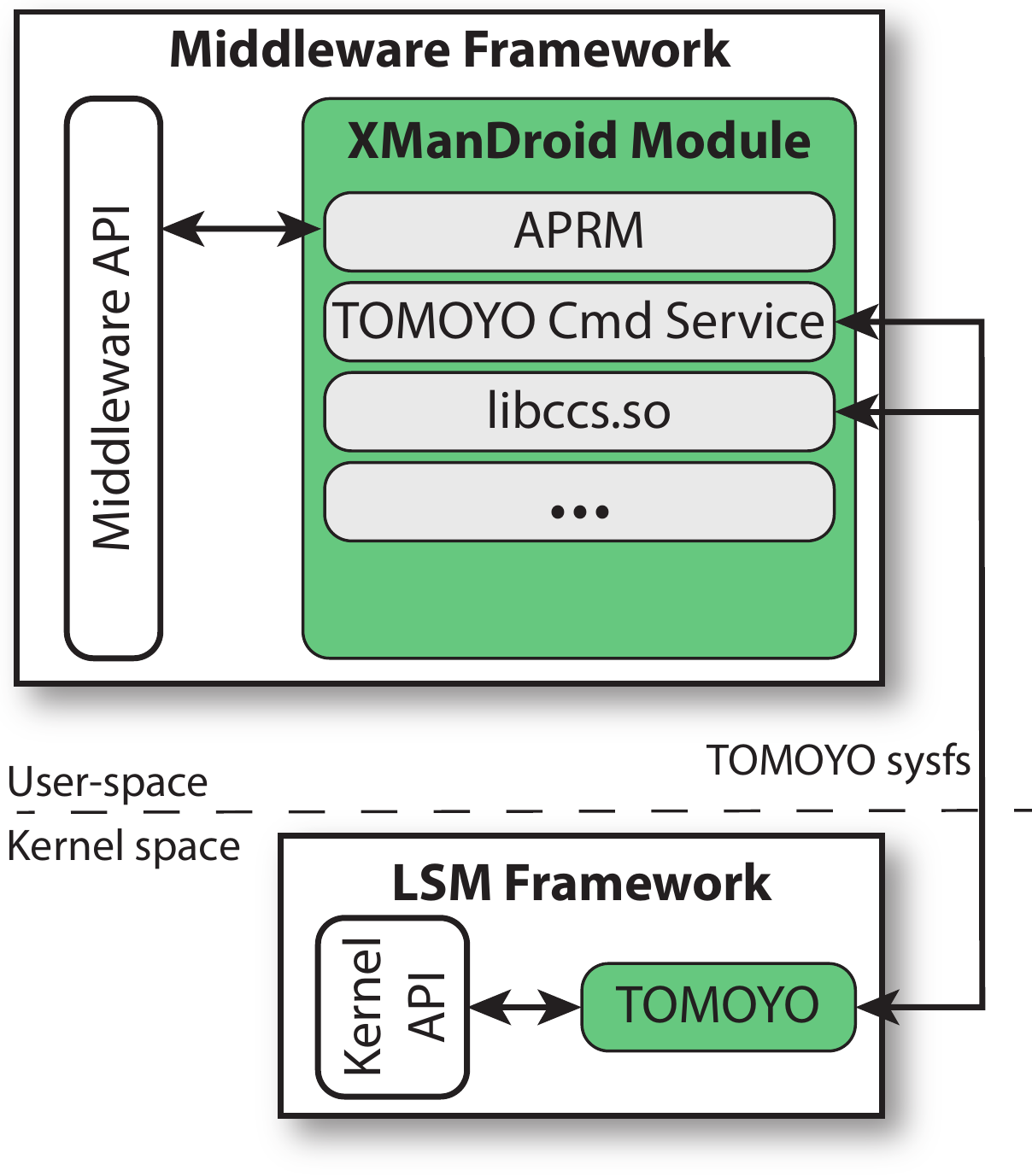}
  \caption{XManDroid module}
  \label{fig:xmandroidmodule}
\end{minipage}
\hfill
\begin{minipage}[b]{.3\linewidth}
  \centering
  \includegraphics[width=.9\linewidth]{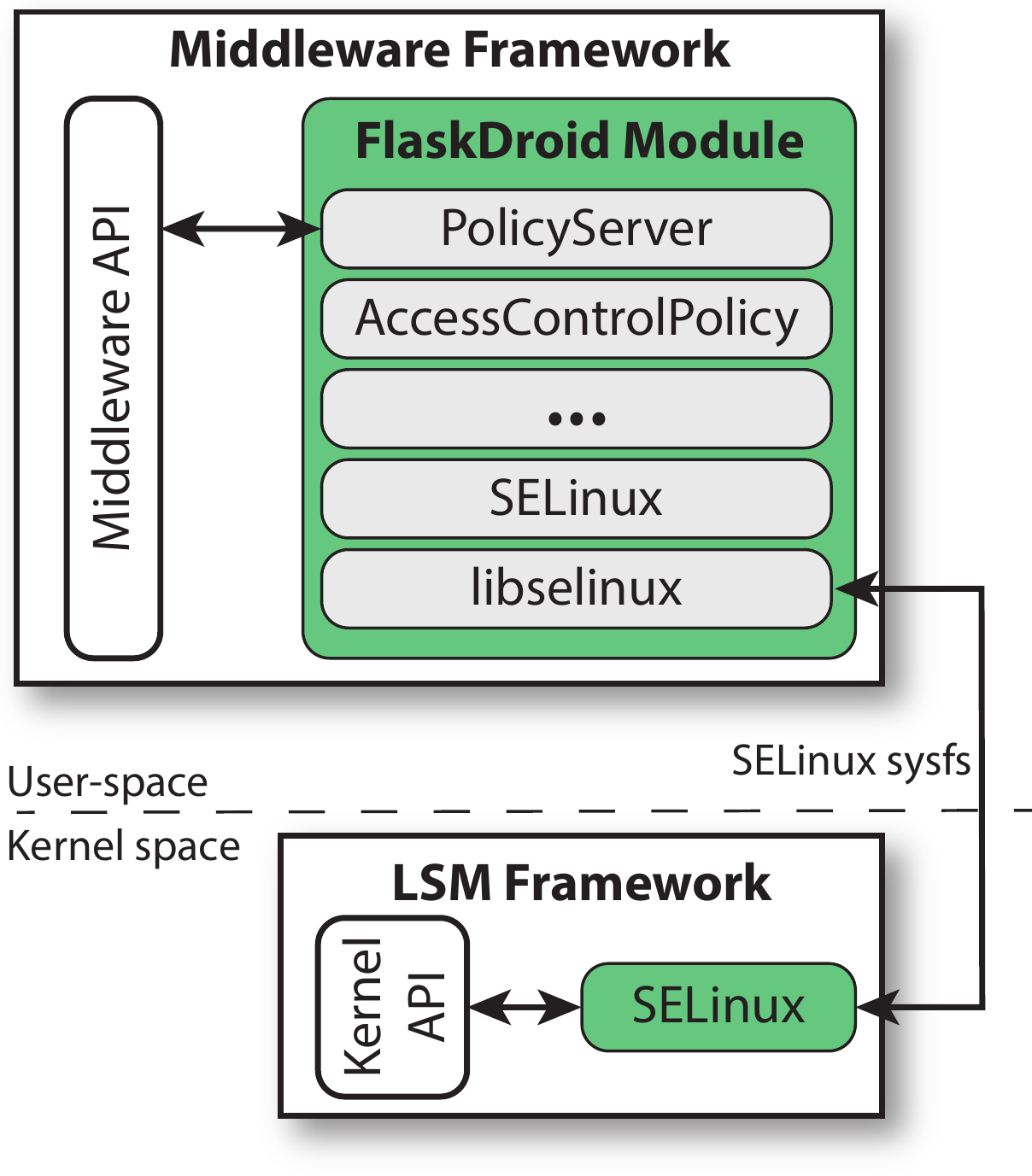}
  \caption{FlaskDroid module}
  \label{fig:flaskdroidmodule}
\end{minipage}
\end{figure*}

\begin{table*}[t]
  \centering
\footnotesize
  \begin{tabular}{|l|r|r|}\hline
    \textbf{Existing solution} & \multicolumn{1}{|c|}{\textbf{LoC of module policy engine}} & \multicolumn{1}{|c|}{\textbf{LoC added/removed/edited (total delta)}} \\\hline
AppGuard~\cite{backes13TACAS} & 5059 & +828/-79/$\circ$\,13~(18.18\%) \\\hline
CRePE~\cite{CoNgCr_10:CRePE} & 3682 & +915/-48/$\circ$\,45~(27.38\%) \\\hline
XManDroid~\cite{BuDaDm_12:TowardsT} & 3244 & +153/-14/$\circ$\,28~(6.01\%) \\\hline
FlaskDroid~\cite{TUD-CS-2013-0115} & 4968 & +749/-32/$\circ$\,40~(16.53\%) \\\hline
  \end{tabular}
  \caption{Effort of porting different security extensions as module on our \OURNAME.}
  \label{tab:usecase:codediff}
\end{table*}

In this section, we demonstrate the efficiency and effectiveness of our \OURNAME by instantiating different security models from related work. To illustrate the versatility of \OURSHORT, we chose models from the areas of inlined reference monitoring, context-based access control, domain isolation, and type enforcement. We present further instantiations of other security models in Appendix~\ref{sec:appendix:furtherusecases}.

\subsection{Inlined Reference Monitoring~\cite{backes13TACAS}}
\label{sec:usecase:appguard}

We use AppGuard~\cite{backes13TACAS} as the use-case to illustrate the applicability of our IRM instrumentation API, but similar application rewriting approaches~\cite{Jeon2012} are also feasible. AppGuard is a privacy app for Android that enables end-users to enforce fine-grained access control policies on 3rd party apps by restricting their ability to access critical system resources. It does so by injecting an IRM into the apps themselves. This approach supports security policies not easily enforceable by traditional external reference monitors in the Android middleware or kernel, e.g., to enforce the use of \emph{https} over \emph{http}.

\paragraph{Implementation as a module:} We ported AppGuard\footnote{Source code provided by the original authors} as a module for \OURSHORT by separating its privacy app into three components:  We adapted the (1)~AppGuard reference monitor with its dynamic hook placement and policy enforcement logic to use the IRM instrumentation API provided by \OURSHORT. The reference monitor is injected into selected app processes via our framework at app startup. The policy decision logic and persistent storage of policy settings was moved into (2)~a middleware module. The middlware module selects the apps into which the IRM is injected. It also implements a Bundle-based communication protocol to exchange policy decisions and security events with the IRM component and with (3)~a front-end app. The front-end app allows the user to adjust policy settings and to view logs of security-relevant events. We used the policies included in the original AppGuard implementation to confirm that policy enforcement by our security module and by the original implementation are identical.

Our AppGuard security module consists of 5059 LoC in total (cf.~Table~\ref{tab:usecase:codediff}), with 782 LoC residing in the middleware module and 4277 LoC in the IRM. Our module diverts in 18.18\% of all LoC from the original code. The majority of the difference, 728 LoC, is attributed to moving the policy decision logic into the middleware module, while only 46 LoC were required to adapt the inlined reference monitor to use the provided instrumentation API.

\subsection{CRePE~\cite{CoNgCr_10:CRePE}}
\label{sec:usecase:crepe}

CRePE is a security extension to Android v2.3 that enforces fine-grained and context-related access control policies. The context is based on the geolocation of the device and, depending on this location, CRePE either allows or denies apps access to security and privacy sensitive information. The security policies can be deployed over different channels, e.g., via SMS. To enforce the policies, CRePE hooked all relevant system services, e.g., to override Android's default permission check with its context-related check.

\paragraph{Implementation as a module:} We ported CRePE\footnote{Source code retrieved from \url{http://sourceforge.net/projects/crepedroid}} as a security module for \OURSHORT by moving its policy engine class \textit{CRePEPolicyManagerService} and related classes, which were originally running as a separate system service, into a module (cf.~Figure~\ref{fig:crepemodule}). On initialization, CRePE's context detector registers as a listener for location updates to detect context changes. Additionally, we used the enforcement functions of our API to re-implement the logic of CRePE's hooks. Furthermore, CRePE uses front-end apps to parse and inject policies from different channels. We moved the policy parser into the module and established a Bundle-based communication protocol between the front-end apps and the module to forward received policies for processing. We used the example policies shipped with the CRePE source code to successfully confirm that the enforcement by our module yields the same results as the original CRePE implementation.

Our port of CRePE as a security module consists of 3682 lines of code (cf.~Table~\ref{tab:usecase:codediff}), excluding the unmodified ANTLR runtime (7526 LoC). Of these 3682 LoC 27.38\% were changed during the port. The bulk of this difference, 817 LoC, is attributed to relocating the policy parser. Implementing the Bundle-based communication protocol added 74 LoC. Only 2 LoC had to be changed to adapt CRePE's calls to the Android API from its original Android v2.3 implementation to our Android v4.3 implementation.

\subsection{XManDroid~\cite{BuDaDm_12:TowardsT}}
\label{sec:usecase:xmandroid}

XManDroid extends the security architecture of Android v2.2.1 to enforce Chinese Wall policies between apps that might jointly leak privacy sensitive information. It uses hooks within different system services in the middleware and TOMOYO Linux at the kernel level to monitor all access control requests, reflect these interactions between processes/apps in a graph model, and use this model to check against policies whether an inter-app communication would lead to an attack state. If so, it denies the new communication. The policy decision logic is implemented as an extension (\textit{APRM}) to the \textit{ActivityManagerService}.

\paragraph{Implementation as a module:} We ported XManDroid\footnote{Source code provided by the original authors} to a module for \OURSHORT (cf.~Figure~\ref{fig:xmandroidmodule}) by extracting the policy decision logic from the \textit{ActivityManagerService} and moving it into a module. Using the enforcement functions of our API we moved the XManDroid hook logic to this module as well and, by using a proprietary channel, we enabled the APRM to communicate from the module with the TOMOYO kernel module. The kernel was specifically compiled and deployed with a TOMOYO Linux security module. The XManDroid source code comes with an example configuration for the policy described in \cite{BuDaDm_12:TowardsT} and we used this configuration to successfully confirm that our module yields the same enforcement results as the original XManDroid implementation.

Our XManDroid middleware module consists of 3244 lines of code, excluding the unchanged JGrapht library (9256 LoC). Our module differs in only 6.01\% (195 LoC) from the original implementation. Of these 195 LoC, 141 are attributed to additions necessary for porting XManDroid's filtering logic for broadcasts from the \textit{ActivityManagerService} to the module.

\subsection{Type Enforcement~\cite{Smalley2013,TUD-CS-2013-0115}}
\label{sec:usecase:flaskdroid}

SE~Android~\cite{Smalley2013} brought SELinux type enforcement to the Android kernel and established the required user space support, e.g., it extended Zygote to label new app processes with a security type. FlaskDroid~\cite{TUD-CS-2013-0115}, developed for Android v4.0.3, extends SE~Android's type enforcement to Android's middleware. Building on SEAndroid's kernel and low-level patches, it adds policy-specific hooks as policy enforcement points to various system services and apps in Android's middleware. The policy decisions at kernel level are made by the SELinux kernel module, while the decisions at middleware are made centrally in a policy server service. Both policy decision points decide based on subject type, object type, and object class reported by the hooks at their respective layer whether control flows should be truncated or not.

\paragraph{Implementation as module:} We realized type enforcement with our \OURSHORT by porting FlaskDroid\footnote{Source code retrieved from \url{http://www.flaskdroid.org/}} as a module (cf.~Figure~\ref{fig:flaskdroidmodule}). At kernel level, we use the SE~Android kernel and provide an SELinux-specific interface implementation for the kernel module. A technical description of this interface implementation is provided in Appendix~\ref{sec:appendix:kmacexample}. Further, we moved the middleware policy server and its dependencies into the middleware module. Using the enforcement functions of our API, we moved the policy-specific hook logic of FlaskDroid into the module as well (cf.~Appendix~\ref{sec:flaskdroiddetail}). Additionally, we used SE~Android's build system to label the file-system with security types.

Our port of FlaskDroid's middleware component as a security module consists of 4968 lines of code (cf.~Table~\ref{tab:usecase:codediff}) and differs in only 16.53\% of all LoC from the original code. The bulk of these changes (550 LoC) is attributed to additions for implementing a mapping from the enforcement functions of our framework API to FlaskDroid's type checks. To confirm the correct enforcement of policies, we used the policies for middleware and kernel level that are provided with the FlaskDroid source code. Additionally, we noticed during our tests that the original implementation contains an error in assigning middlware security types to processes.\footnote{It maps process UIDs always to the security type of the first package in a shared sandbox, although the policy can define different types for packages that share a UID.} Additional changes were necessary to fix this error in our FlaskDroid module.

\section{Evaluation and Discussion}
\label{sec:eval}
In this section we evaluate the performance of our \OURNAME and discuss its current scope and prospective future work.

\begin{figure}[t]
  \centering
  \includegraphics[width=\linewidth]{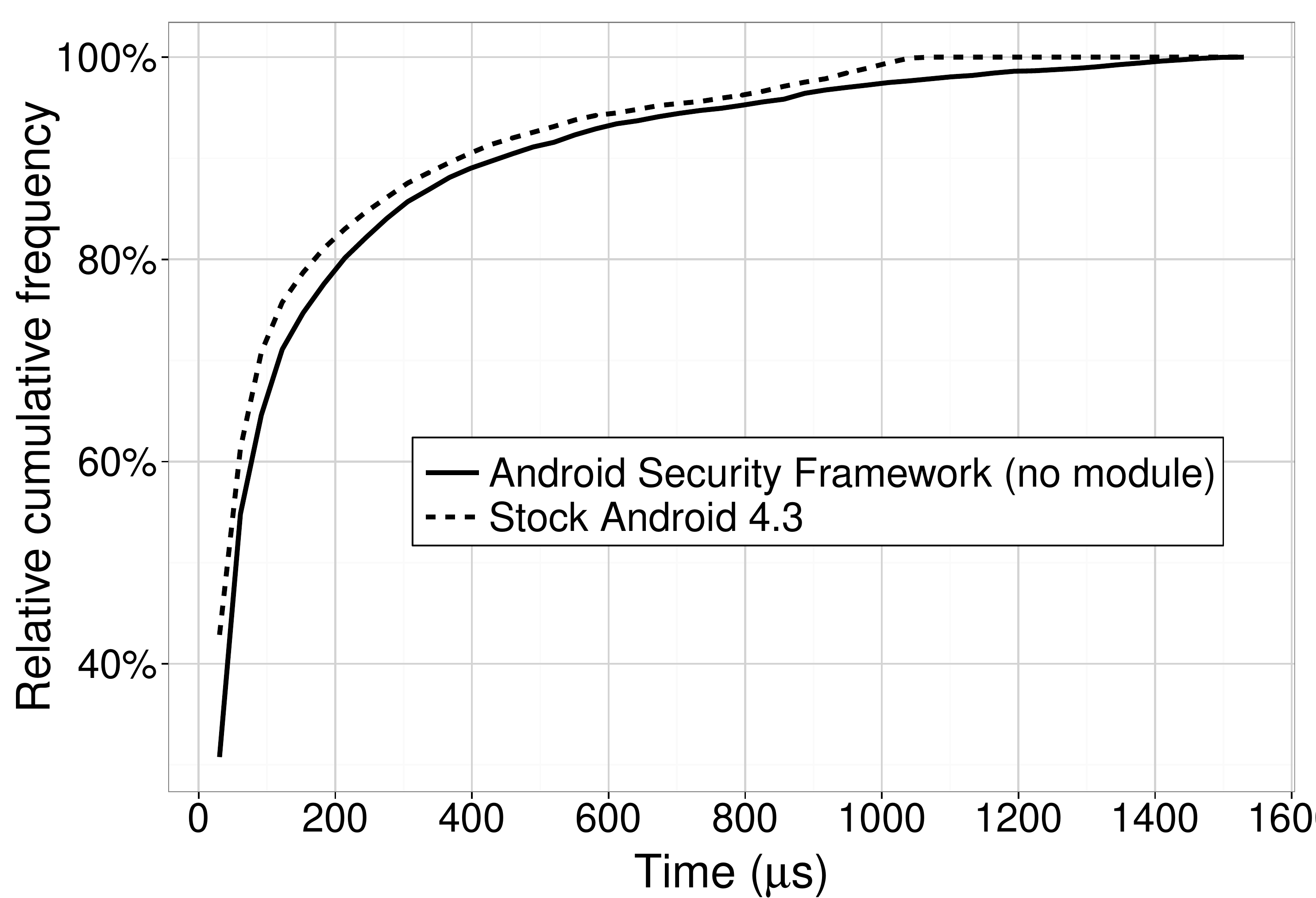}
  \caption{Relative cumulative frequency distribution of micro benchmarks in stock Android (dashed line) vs.~\OURNAME (solid line).}
  \label{fig:eval:cumulative}
\end{figure}

\begin{table}[t]
  \centering
  \small
  \begin{tabular}{|p{3cm}|r|l|}\cline{2-3}
    \multicolumn{1}{c|}{} & \textbf{Frequency} & \textbf{Mean} ($\mu$s)\\\cline{1-3}
    \textbf{Stock Android~4.3} & 7320 & 116.182$\pm$4.550\\\hline
\textbf{\OURSHORT v4.3} & 6009 & 129.851$\pm$5.681\\\hline
  \end{tabular}
  \caption{Weighted average performance overhead of executing
    hooked functions in stock Android and in our \OURNAME. The
    margin of error is given for the 95\% confidence interval.}
  \label{table:eval:averageperformance}
\end{table}

\subsection{Performance}
\label{sec:eval:performance}

Although the actual performance overhead strongly depends on the overhead imposed by the loaded module, we wanted to establish a baseline for the impact of our \OURNAME on the system performance. The performance of LSM has been evaluated separately, e.g., for SEAndroid~\cite{Smalley2013}, and we are interested here in the effect of our new middleware security framework on the performance of instrumented middleware system services and apps.

\paragraph{Methodology.}
We implemented our \OURSHORT as a modification to the Android OS code base in version 4.3\_r3.1 (\textit{``Jelly Bean''}) and used the Android Linux kernel in branch \textit{android-omap-tuna-3.0-jb-mr1.1}. We performed micro-benchmarks for all execution paths on which a hook diverts the control flow to our middleware framework: We first measured the execution time of each hooked function with no security module loaded and allowing by default all access. Afterwards we repeated this test with hooks disabled to measure the default performance of the same functions and thus operating like a stock Android. All our micro-benchmarks were performed on a standard Nexus 7 development tablet (Quad-core 1.51 GHz CPU and 2GB DDR3L RAM), which we booted and then used according to a testplan for different daily tasks such as browsing the Internet, sending text messages and e-mails, contacts management, or (un-)installing 3rd party apps.

\paragraph{Micro-benchmark results.}
Table~\ref{table:eval:averageperformance} presents the number of measurements for each test case and their mean values. To exclude extreme outliers, we excluded in both measurement series the highest decile of the measurements. For \OURSHORT the mean is the weighted mean value with consideration of the frequency of each single hook. Table~\ref{tab:eval:tophooks} in Appendix~\ref{sec:appendix:hookfreq} provides a break down of the most frequently called hooked Android API functions and their mean execution time. In overall, our framework with no loaded module imposed with 129.851~$\mu$s approximately only 11.8\% overhead compared to stock Android. Figure~\ref{fig:eval:cumulative} presents the relative cumulative frequency distribution of our measurements series and further illustrates this low performance overhead. Appendix~\ref{sec:appendix:moduleperf} provides an overview of the micro-benchmark results for our example modules from Section~\ref{sec:usecases}.

\subsection{Current Scope and Future Work}
\label{sec:eval:limitations}

\paragraph{System setup.}

Certain security models require a preparatory system setup. For instance, type enforcement requires a pre-labelling of all subjects and objects as well as enabling the SELinux kernel module. After the system has been setup, \OURSHORT supports modularization of these security models (cf.~Section~\ref{sec:usecase:flaskdroid}).

\paragraph{Module Integrity.}

As part of the kernel, the \KMODULE has the highest level of integrity. In contrast, the \MMODULE, as a user space process, can be circumvented or compromised by attacks against the underlying system (e.g., root exploits) and thus requires support by the kernel modules to prevent low-level privilege escalation attacks. Inlined reference monitors are inherently susceptible to attacks by malicious applications, because the reference monitor executes in the same process as the application that it monitors and no strong security boundary exists between the monitor and the app code. To remedy this situation, we are currently retrofitting Android's application model to combine the benefits of inlined and of system-centric reference monitors. By splitting apps into smaller units of trust (e.g., app components and ad libs), system-centric reference monitors are able to differentiate distinct trust levels within apps~\cite{Provos:2003:PPE:1251353.1251369,Wang:2014:compasec,Shekhar:2012:ASS:2362793.2362821,Pearce:2012:APS:2414456.2414498}.

\paragraph{Completeness.}
It is crucial for the effectiveness of our security framework, that \textit{all} access to security and privacy sensitive resources is mediated by the reference monitors. We consider it out of scope for this submission to formally verify the completeness of our prototype framework, but plan to use recent advances in static and dynamic analysis on Android to verify the placement of our hooks, similarly to how it was done for the LSM framework~\cite{edwards:ccs02,ganapathy:ccs05}.

\paragraph{Information flow control.}
Our framework provides modules with the control over which subject (e.g., app) has \textit{access to} which objects (e.g., device location), but it cannot control how privileged subjects \textit{distribute} this information. Controlling information flows is an orthogonal problem specifically addressed by solutions such as \textit{TaintDroid}~\cite{EnGiBy_10:Taindroid},\textit{AppFence}~\cite{HoHaJuScWe_11:RetrofittingAndroid}, or \textit{MOSES}~\cite{Russello:2012:MSO:2295136.2295140}. We plan to integrate such data flow solutions into our framework and to extend our security API with new generic calls for taint labeling and taint checking.

\paragraph{Framework Maintenance.}
\OURSHORT removes the burden for authors of security extensions to patch every Android OS version. Instead it puts this burden onto the framework maintainer, which is preferably the OS vendor like Google. In light of the vendor's expert knowledge about the Android software stack, we consider the maintenance overhead for integrating \OURSHORT into the continuous integration of the Android development as moderate. In particular the recent development history of Android~v4.x, which had one focus clearly set on security (e.g., integration of SELinux, use of capabilities, AppOps and IntentFirewall), makes us optimistic that the required commitment by the vendor is viable.

\section{Conclusion}
\label{conclusion}

In this paper we presented the \OURNAME~(\OURSHORT), an extensible and policy-agnostic security infrastructure for Android. \OURSHORT allows security experts to develop Android security extensions against a novel Android security API and to deploy their solutions in form of modules or ``security apps''. Modularizing security extensions overcomes the current unsatisfactory situation that policy authors are either limited to one predetermined security model that is embedded in the Android software stack or that they are forced to confide in a security-model-specific Android fork instead of the mainline Android code base. Additionally, this modularization provides a number of benefits such as easier maintenance and direct comparison of security extensions. We demonstrated the effectiveness and efficiency of \OURSHORT by porting different security models from related work to \OURSHORT modules and by establishing a baseline for the impact of our infrastructure on the system performance.

The source code of our \OURNAME and our example modules can be anonymously retrieved from \url{http://infsec.cs.uni-saarland.de/projects/asf/ASF_code.zip}.

{\footnotesize \bibliographystyle{acm}
\bibliography{asf}}

\appendix

\onecolumn

\section*{Appendix}

\section{Policy Module Interface}
\label{sec:appendix:api}

\begin{lstlisting}[emph={},basicstyle=\footnotesize,caption={Interface for Access Control Policy Modules}]
public interface IAccessControlModule {
  /* ***********************************************
   * General functions
   *********************************************** */
  public boolean init();
  public ModuleConfiguration getConfig();
  public void shutdown();
  
  /* ***********************************************
   *  Package life-cycle event hooks
   *********************************************** */
  public void security_event_installNewPackage(PackageParser.Package pkg, UserHandle user);
  public void security_event_replacePackage(PackageParser.Package oldPkg, PackageParser.Package newPkg, UserHandle user);
  public void security_event_deletePackage(String packageName, int uid, int removedAppId, int removedUsers[], UserHandle user);
  
  /* ***********************************************
   *  Generic hooks
   *********************************************** */
  public void security_generic_checkPolicy(Bundle arguments);
  public void security_generic_callModule(Bundle arguments);
  public boolean security_generic_instrumentApp(String packageName);

  /* ***********************************************
   *  Broadcast hooks
   *********************************************** */
  public boolean security_broadcast_deliverToRegisteredReceiver(Intent intent, ComponentName targetComp, String requiredPermission, int targetUid, int targetPid, String callerPackage, ApplicationInfo callerApp, int callingUid, int callingPid);
  public boolean security_broadcast_processNextBroadcast(Intent intent, ResolveInfo target, String requiredPermission, String callerPackage, ApplicationInfo callerApp, int callingUid, int callingPid);
  
  /* ***********************************************
   *  ContentProvider.Transport hooks
   *********************************************** */
  public boolean security_cp_applyOperation(ContentProviderOperation op, int uid, int pid);
  public boolean security_cp_preQuery(String callingPkg, Uri uri, String[] projection, String selection, String[] selectionArgs, String sortOrder, int uid, int pid);
  public Cursor security_cp_postQuery(Cursor result, String callingPkg, Uri uri, String[] projection, String selection, String[] selectionArgs, String sortOrder, int uid, int pid);
  public boolean security_cp_insert(Uri uri, ContentValues initialValues, int uid, int pid);
  public boolean security_cp_bulkInsert(Uri uri, ContentValues[] initialValues, int uid, int pid);
  public boolean security_cp_delete(String callingPkg, Uri uri, String selection, String[] selectionArgs, int uid, int pid);
  public boolean security_cp_update(String callingPkg, Uri uri, ContentValues values, String selection, String[] selectionArgs, int uid, int pid);
  public boolean security_cp_openFile(Uri uri, String mode, int uid, int pid);
  public boolean security_cp_preCall(String providerClass, String method, String arg, Bundle extras, int uid, int pid);
  public Bundle security_cp_postCall(Bundle result, String providerClass, String method, String arg, Bundle extras, int uid, int pid);
  
  public boolean security_contacts_preQueryDirectory(Uri uri, String directoryName, String directoryType, String[] projection, String selection, String[] selectionArgs, String sortOrder, int uid, int pid);
  public BulkCursorDescriptor security_contacts_postQueryDirectory(BulkCursorDescriptor result, String directoryName, String directoryType, String providerName, Uri uri, String[] projection, String selection, String[] selectionArgs, String sortOrder, int uid, int pid);
  
  /* ***********************************************
   *  Activity related hooks
   *********************************************** */
  public boolean security_ams_startActivity(Intent intent, String resolvedType, ActivityInfo aInfo, String resultWho, int requestCode, int startFlags, Bundle options, ApplicationInfo callerInfo, int callingPid, int callingUid, int callingUserId);
  public boolean security_ams_finishActivity(ComponentName origActivity, ComponentName realActivity, Intent intent, int userId, ApplicationInfo info, int resultCode, Intent resultData, int uid, int pid);
  public boolean security_ams_moveTaskToFront(ComponentName origActivity, ComponentName realActivity, Intent intent, int userId, ApplicationInfo info, int flags, Bundle options, int uid, int pid);
  public boolean security_ams_moveTaskToBack(ComponentName origActivity, ComponentName realActivity, Intent intent, int userId, ApplicationInfo info, int uid, int pid);
  public boolean security_ams_clearApplicationUserData(String packageName, int pkgUid, int userId, int uid, int pid);
  
  /* ***********************************************
   * Permission check overrides
   *********************************************** */
  public int security_ams_checkComponentPermission(String permission, int origUid, int origPid, int tlsUid, int tlsPid, int owningUid, boolean exported, int callerUid, int callerPid);
  public boolean security_ams_checkCPUriPermission(Uri uri, ProviderInfo cpi, int processUid, int processPid, boolean procesIsolated, int processUserId, String processName, ApplicationInfo info, int uid, int pid);
  public boolean security_ams_checkCPUriPermission(Uri uri, ProviderInfo cpi, int uid, int pid);
  public boolean security_ams_checkGrantUriPermission(int callingUid, String targetPkg, int targetUid, Uri uri, int modeFlags);
  public int security_ams_checkUriPermission(Uri uri, int origUid, int origPid, int tlsUid, int tlsPid, int modeFlags);
  
  /* ***********************************************
   * PackageManagerService hooks
   *********************************************** */
  public boolean security_pms_getPackageInfo(PackageInfo pi, int flags, int userId, boolean isUninstalled, int uid, int pid);
  public boolean security_pms_getPackageUid(ApplicationInfo info, int userId, int uid, int pid);
  public boolean security_pms_getPackageGids(ApplicationInfo info, int[] gids, int uid, int pid);
  public String[] security_pms_getPackagesForUid(int forUid, String[] packages, int uid, int pid);
  public boolean security_pms_getNameForUid(int forUid, String name, int uid, int pid);
  public boolean security_pms_getUidForSharedUser(String sharedUserName, int suid, int uid, int pid);
  public boolean security_pms_findPreferredActivity(Intent intent, String resolvedType, int flags, ResolveInfo ri, int priority, int userId, int uid, int pid);
  public List<ResolveInfo> security_pms_queryIntentActivities(List<ResolveInfo> currentList, Intent intent, String resolvedType, int flags, int userId, int uid, int pid);
  public List<ResolveInfo> security_pms_queryIntentReceivers(List<ResolveInfo> currentList, Intent intent, String resolvedType, int flags, int userId, int uid, int pid);
  public List<ResolveInfo> security_pms_queryIntentServices(List<ResolveInfo> currentList, Intent intent, String resolvedType, int flags, int userId, int uid, int pid);
  public ArrayList<PackageInfo> security_pms_getInstalledPackages(ArrayList<PackageInfo> currentList, int flags, int userId, int uid, int pid);
  public ArrayList<PackageInfo> security_pms_getPackagesHoldingPermissions(ArrayList<PackageInfo> currentList, int flags, int userId, String[] permissions, int uid, int pid);
  public ArrayList<ApplicationInfo> security_pms_getInstalledApplications(ArrayList<ApplicationInfo> currentList, int flags, int userId, int uid, int pid);
  public ArrayList<ApplicationInfo> security_pms_getPersistentApplications(ArrayList<ApplicationInfo> currentList, int flags, int uid, int pid);
  public boolean security_pms_getProviderInfo(ProviderInfo pi, ComponentName component, int flags, int userId, int uid, int pid);
  public boolean security_pms_getActivityInfo(ActivityInfo ai, ComponentName component, int flags, int userId, int uid, int pid);
  public boolean security_pms_getReceiverInfo(ActivityInfo ai, ComponentName component, int flags, int userId, int uid, int pid);
  public boolean security_pms_getServiceInfo(ServiceInfo si, ComponentName component, int flags, int userId, int uid, int pid);
  /* Pre-init function (packages are scanned before init is called) */
  public boolean security_pms_scanPackage(PackageParser.Package pkg);
  public boolean security_pms_deletePackage(PackageParser.Package pkg, boolean isSystemApp, boolean dataOnly, int flags);
  public boolean security_pms_deletePackageSingleUser(PackageParser.Package pkg, boolean isSystemApp, int flags, int user);
  
  /* ***********************************************
   * Content Provider (general) related hooks 
   *********************************************** */
  /* Changed ProcessRecord to public for our module SDK */
  public boolean security_ams_checkContentProviderPermission(ProviderInfo cpi, String permission, int processUid, int processPid, boolean procesIsolated, int processUserId, String processName, ApplicationInfo info, int uid, int pid);
  public boolean security_ams_checkContentProviderPermission(ProviderInfo cpi, String permission, int uid, int pid);
  public boolean security_ams_checkPathPermission(ProviderInfo cpi, PathPermission pp, String permission, int processUid, int processPid, boolean procesIsolated, int processUserId, String processName, ApplicationInfo info, int uid, int pid);
  public boolean security_ams_checkPathPermission(ProviderInfo cpi, PathPermission pp, String permission, int uid, int pid);
  public boolean security_ams_checkAppSwitchAllowed(int uid, int pid);
  
  /* ***********************************************
   * Service related hooks
   *********************************************** */
  public List<ActivityManager.RunningServiceInfo> security_ams_getServices(ArrayList<ActivityManager.RunningServiceInfo> srvList, int uid, int pid);
  public boolean security_ams_peekService(Intent service, String resolvedType, ServiceInfo serviceInfo, ApplicationInfo appInfo, String packageName, String permission, int uid, int pid);
  public boolean security_ams_startService(Intent service, String resolvedType, ComponentName name, String shortName, ServiceInfo serviceInfo, ApplicationInfo appInfo, int srvUserId, String packageName, String processName, String permission, int callingPid, int callingUid);
  public boolean security_ams_stopService(Intent service, String resolvedType, ComponentName name, String shortName, ServiceInfo serviceInfo, ApplicationInfo appInfo, int srvUserId, String packageName, String processName, String permission, int callingPid, int callingUid);
  public boolean security_ams_bindService(Intent service, String resolvedType, int flags, ComponentName name, String shortName, ServiceInfo serviceInfo, ApplicationInfo appInfo, int srvUserId, String packageName, String processName, String permission, int callingPid, int callingUid);
  
  /* ***********************************************
   * LocationManagerService hooks
   *********************************************** */
  public void security_location_getAllProviders(List<String> providerList, int uid, int pid);
  public void security_location_getProviders(List<String> providers, Criteria criteria, boolean enabledOnly, int uid, int pid);
  /* Unhide LocationRequest for our module SDK */
  public void security_location_requestLocationUpdates(LocationRequest request, PendingIntent pi, int uid, int pid);
  public void security_location_removeLocationUpdates(PendingIntent pi, int uid, int pid);
  public Location security_location_getLastLocation(Location currentLocation, LocationRequest request, int uid, int pid);
  public boolean security_location_addGpsStatusListener(int uid, int pid);
  public boolean security_location_sendExtraCommand(String provider, String command, Bundle extras, int uid, int pid);
  /* Unhide Geofence class for our module SDK */
  public void security_location_requestGeofence(LocationRequest request, Geofence geofence, PendingIntent intent, int uid, int pid);
  public void security_location_removeGeofence(Geofence geofence, PendingIntent intent, int uid, int pid);
  public boolean security_location_isProviderEnabled(String provider, int uid, int pid);
  public Location security_location_reportLocation(Location location, boolean passive, int uid, int pid);
  public ProviderProperties security_location_addTestProvider(String name, ProviderProperties properties, int uid, int pid);
  public boolean security_location_removeTestProvider(String provider, int uid, int pid);
  public boolean security_location_setTestProviderLocation(String provider, Location location, int uid, int pid);
  public boolean security_location_clearTestProviderLocation(String provider, int uid, int pid);
  public boolean security_location_setTestProviderEnabled(String provider, boolean enabled, int uid, int pid);
  public boolean security_location_clearTestProviderEnabled(String provider, int uid, int pid);
  public boolean security_location_setTestProviderStatus(String provider, int status, Bundle extras, long updateTime, int uid, int pid);
  public boolean security_location_clearTestProviderStatus(String provider, int uid, int pid);
  public boolean security_location_sendLocationUpdate(Location location, String receiverPackageName, int pid, int uid);
  public boolean security_location_updateFence(Location location, Geofence fence, PendingIntent fenceIntent, String fencePackageName, int uid);
  
  /* ***********************************************
   * AudioService hooks
   *********************************************** */
  public boolean security_audio_adjustStreamVolume(int streamType, int direction, int flags, int uid, int pid);
  public boolean security_audio_setStreamVolume(int streamType, int index, int flags, int uid, int pid);
  public boolean security_audio_setMasterVolume(int volume, int flags, int uid, int pid);
  public boolean security_audio_setRingerMode(int mode, int uid, int pid);
  public boolean security_audio_setSpeakerphoneOn(boolean on, int uid, int pid);
  
  /* ***********************************************
   * TelephonyService hooks
   *********************************************** */
  public boolean security_telephony_call(String number, int uid, int pid);
  public List<NeighboringCellInfo> security_telephony_getNeighboringCellInfo(List<NeighboringCellInfo> currentList, int uid, int pid);
  
  /* ***********************************************
   * SMS and MMS Service hooks
   *********************************************** */
  public boolean security_sms_copyMessageToIcc(int status, byte[] pdu, byte[] smsc, int uid, int pid);
  public boolean security_sms_getAllMessagesFromIcc(int uid, int pid);
  public List<RawByteData> security_sms_getAllMessagesFromIccFilter(List<RawByteData> rawSms, int uid, int pid);
  public boolean security_sms_sendData(String destAddr, String scAddr, int destPort, byte[] data, PendingIntent sentIntent, PendingIntent deliveryIntent, int uid, int pid);
  public boolean security_sms_sendText(String destAddr, String scAddr, String text, PendingIntent sentIntent, PendingIntent deliveryIntent, int uid, int pid);
  public boolean security_sms_sendMultipartText(String destAddr, String scAddr, List<String> parts, List<PendingIntent> sentIntents, List<PendingIntent> deliveryIntents, int uid, int pid);
  public boolean security_sms_updateMessageOnIccEf(int index, int status, byte[] pdu, int uid, int pid);
  
  /* ***********************************************
   * WiFi Service hooks
   *********************************************** */
  public List<ScanResult> security_wifi_getScanResult(List<ScanResult> result, int uid, int pid);
  public boolean security_wifi_startScan(int uid, int pid);
  public boolean security_wifi_setWifiEnabled(boolean enable, int uid, int pid);
  public boolean security_wifi_setWifiApEnabled(WifiConfiguration wifiConfig, boolean enabled, int uid, int pid);
  public boolean security_wifi_setWifiApConfiguration(WifiConfiguration wifiConfig, int uid, int pid);
  public boolean security_wifi_disconnect(int uid, int pid);
  public boolean security_wifi_reconnect(int uid, int pid);
  public boolean security_wifi_reassociate(int uid, int pid);
  public List<WifiConfiguration> security_wifi_getConfiguredNetworks(List<WifiConfiguration> currentList, int uid, int pid);
  public boolean security_wifi_addOrUpdateNetwork(WifiConfiguration config, int uid, int pid);
  public boolean security_wifi_removeNetwork(int netId, int uid, int pid);
  public boolean security_wifi_enableNetwork(int netId, boolean disableOthers, int uid, int pid);
  public boolean security_wifi_disableNetwork(int netId, int uid, int pid);
  public boolean security_wifi_getConnectionInfo(WifiInfo info, int uid, int pid);
  public boolean security_wifi_setCountryCode(String countryCode, boolean persist, int uid, int pid);
  public boolean security_wifi_setFrequencyBand(int band, boolean persist, int uid, int pid);
  public boolean security_wifi_startWifi(int uid, int pid);
  public boolean security_wifi_stopWifi(int uid, int pid);
  public boolean security_wifi_addToBlacklist(String bssid, int uid, int pid);
  public boolean security_wifi_clearBlacklist(int uid, int pid);
  public boolean security_wifi_getWifiServiceMessenger(int uid, int pid);
  public boolean security_wifi_getWifiStateMachineMessenger(int uid, int pid);
  public boolean security_wifi_getConfigFile(String currentConfig, int uid, int pid);
  
  /* ***********************************************
   * ClipboardService hooks
   *********************************************** */
  public ClipData security_clip_getPrimaryClip(ClipData currentPrimary, int clipUid, int uid, int pid);
  public boolean security_clip_setPrimaryClip(ClipData clip, int uid, int pid);
  public boolean security_clip_informPrimaryClipChanged(ClipData currentPrimary, int setByUid, String packageName, int uid);
  public ClipDescription security_clip_getPrimaryClipDescription(ClipDescription currentDescription, int clipUid, int uid, int pid);
  public boolean security_clip_hasPrimaryClip(boolean hasClipboard, int clipUid, int uid, int pid);
  public boolean security_clip_hasClipboardText(String currentText, int clipUid, int uid, int pid);
  
  /* ***********************************************
   * PowerManagerService hooks
   *********************************************** */
  public boolean security_power_acquireWakeLock(String tag, WorkSource ws, int uid, int pid);
  public boolean security_power_userActivity(long eventTime, int event, int flags, int uid, int pid);
  public boolean security_power_goToSleep(long eventTime, int reason, int uid, int pid);
  public boolean security_power_wakeUp(long eventTime, int uid, int pid);
  public boolean security_power_nap(long time, int uid, int pid);
  public boolean security_power_setBacklightBrightness(int brightness, int uid, int pid);
  public boolean security_power_reboot(boolean confirm, String reason, boolean wait, int uid, int pid);
  
  /* ***********************************************
   * PhoneSubscriberInfo hooks
   *********************************************** */
  public String security_phonesubinfo_getDeviceId(String id, int uid, int pid);
  public String security_phonesubinfo_getDeviceSvn(String svn, int uid, int pid);
  public String security_phonesubinfo_getSubscriberId(String id, int uid, int pid);
  public String security_phonesubinfo_getGroupIdLevel1(String groupid, int uid, int pid);
  public String security_phonesubinfo_getIccSerialNumber(String icc, int uid, int pid);
  public String security_phonesubinfo_getLine1Number(String number, int uid, int pid);
  public String security_phonesubinfo_getLine1AlphaTag(String tag, int uid, int pid);
  public String security_phonesubinfo_getMsisdn(String msisdn, int uid, int pid);
  public String security_phonesubinfo_getVoiceMailNumber(String number, int uid, int pid);
  public String security_phonesubinfo_getVoiceMailAphaTag(String tag, int uid, int pid);
  public String security_phonesubinfo_getIsimImpi(String impi, int uid, int pid);
  public String security_phonesubinfo_getIsimDomain(String domain, int uid, int pid);
  public String[] security_phonesubinfo_getIsimImpu(String impu[], int uid, int pid);
}
\end{lstlisting}

\begin{lstlisting}[language=Java,basicstyle=\footnotesize,caption={Interface for Access Control Policy Modules to Linux Security Module},label={listing:kmacapi}]
public interface KMACAdaptor {
  public boolean init();
  public boolean isEnabled();
  public boolean isEnforcing();
  public boolean setEnforcing(boolean value);
  public boolean setContext(String path, Bundle context);
  public boolean restoreContext(Bundle context);
  public Bundle getContext(String path);
  public Bundle getPeerContext(FileDescriptor fd); /* wrapper around getsockopt call to LSM */
  public Bundle getCurrentContext();
  public Bundle getProcessContext(int pid);
  public Bundle getConfig(Bundle args); /* e.g., get list of defined booleans or one specific boolean value */
  public boolean setConfig(Bundle conf); /* e.g., set a boolean value */
  public boolean checkAccess(Bundle args); /* args can be, e.g., quadruple of subject ctx, object ctx, object class, op */
  
  /* Zygote is statically integrated with the Kernel MAC, thus, each KMACAdaptor must implemented these hooks in ZygoteConnection */
  public boolean security_zygote_applyUidSecurityPolicy(Credentials creds, Bundle peerSecurityContext);
  public boolean security_zygote_applyRlimitSecurityPolicy(Credentials creds, Bundle peerSecurityContext);
  public boolean security_zygote_applyCapabilitiesSecurityPolicy(Credentials creds, Bundle peerSecurityContext);
  public boolean security_zygote_applyInvokeWithSecurityPolicy(Credentials creds, Bundle peerSecurityContext);
  public boolean security_zygote_applySecurityLabelPolicy(Credentials creds, Bundle peerSecurityContext);
}
\end{lstlisting}

\begin{lstlisting}[language=Java,basicstyle=\footnotesize,caption={Methods for IRM instrumentation},label={listing:instrumentation}]
public class Instrumentation {
  public static void initClass(Class<?> clazz);

  public static int redirectMethod(String fromDescriptor, String toDescriptor);
  public static int redirectMethod(Signature from, Signature to);

  public static void callVoidMethod(Class<?> caller, Object _this, Object... args);
  public static void callVoidMethod(String id, Object _this, Object... args);
  public static void callVoidMethod(int methodId, Object _this, Object... args);
  public static int callIntMethod(Class<?> caller, Object _this, Object... args);
  public static int callIntMethod(String id, Object _this, Object... args);
  public static int callIntMethod(int methodId, Object _this, Object... args);
  public static boolean callBooleanMethod(Class<?> caller, Object _this, Object... args);
  public static boolean callBooleanMethod(String id, Object _this, Object... args);
  public static boolean callBooleanMethod(int methodId, Object _this, Object... args);
  public static Object callObjectMethod(Class<?> caller, Object _this, Object... args);
  public static Object callObjectMethod(String id, Object _this, Object... args);
  public static Object callObjectMethod(int methodId, Object _this, Object... args);
  public static void callStaticVoidMethod(Class<?> caller, Class<?> _clazz, Object... args);
  public static void callStaticVoidMethod(String id, Class<?> _clazz, Object... args);
  public static void callStaticVoidMethod(int methodId, Class<?> _clazz, Object... args);
  public static Object callStaticObjectMethod(Class<?> caller, Class<?> _clazz, Object... args);
  public static Object callStaticObjectMethod(String id, Class<?> _clazz, Object... args);
  public static Object callStaticObjectMethod(int methodId, Class<?> _clazz, Object... args);
}
\end{lstlisting}


\section{Break Down of Policy Enforcement Coverage}
\label{sec:appendix:coverage}

\begin{table}[h]
  \centering
  \small
  \begin{tabular}{|l|r|p{8cm}|}\hline
    \textbf{System App/Service} & \textbf{Number of hooks} & \multicolumn{1}{|c|}{\textbf{Example hooks}}\\\hline\hline
    BroadcastQueue & 2 & deliverToRegisteredReceiver, processNextBroadcast \\\hline
    ContentProvider & 12 & insert, update, preQuery, postQuery \\\hline
    ActivityStack & 5 & startActivity, moveTaskToBack, finishActivity \\\hline
    ActivityManagerService & 10 & checkComponentPermission, checkUriPermission, checkGrantUriPermission\\\hline
    PackageManagerService & 21 & getPackageInfo, findPreferredActivity, queryIntentReceivers, getServiceInfo, scanPackage, deletePackage\\\hline
ActiveServices & 5 & startService, bindService, getServices\\\hline
LocationManagerService & 21 & getProviders, requestLocationUpdates, requestGeofence, reportLocation, setTestProviderLocation\\\hline
AudioService & 5 & adjustStreamVolume, setMasterVolume, setRingerMode\\\hline
TelephonyService & 2 & call, getNeighboringCells\\\hline
SMSService & 7 & getAllMessagesFromIcc, sendData, sendText\\\hline
WiFiService & 23 & getScanResults, addOrUpdateNetwork, getConnectionInfo, getWifiServiceManager\\\hline
ClipboardService & 7 & getPrimaryClip, setPrimaryClip\\\hline
PowerManagerService & 6 & acquireWakeLock, userActivity, reboot\\\hline
PhoneSubInfo & 13 & getDeviceId, getIccSerialNumber, getLine1Number, getIsimImpi\\\hline
\multicolumn{1}{c|}{} & \multicolumn{1}{r|}{\textbf{Total:\hfill139}} & \multicolumn{1}{c}{} \\\cline{2-2}
  \end{tabular}
  \caption{Break down of hooked system apps and services.}
  \label{tab:coveragebreakdown}
\end{table}

\section{Break Down of Most Frequently Called Enforcement Hooks}
\label{sec:appendix:hookfreq}

\begin{table*}[h]
  \centering
  \footnotesize
  \begin{tabular}{|l|r|r|r|r|}\cline{2-5}
    \multicolumn{1}{l}{} & \multicolumn{2}{|c|}{\textbf{Android Security Framework}} & \multicolumn{2}{c|}{\textbf{Stock Android v4.3}}\\\hline
    \textbf{Hooked function} & \textbf{Frequency} & \textbf{Mean ($\mu$s)}  & \textbf{Frequency} & \textbf{Mean ($\mu$s)}\\\hline\hline
    ActivityManagerService.checkComponentPermission & 1705 & 39.413$\pm$0.658 & 2024 & 36.518$\pm$0.523\\\hline
    BroadcastQueue.broadcastIntent & 908 & 305.274$\pm$16.752 & 1007 & 332.328$\pm$17.085\\\hline
    SettingsProvider.call & 544 & 67.710$\pm$3.004 & 669 & 46.574$\pm$1.723\\\hline
    PackageManagerService.queryIntentReceivers & 438 & 92.458$\pm$3.598 & 745 & 84.343$\pm$2.296\\\hline
    PackageManagerService.queryIntentActivities & 296 & 192.178$\pm$15.458 & 242 & 195.211$\pm$18.355\\\hline
    PowerManagerService.acquireWakeLock & 229 & 296.246$\pm$10.740 & 255 & 295.601$\pm$11.121\\\hline
    PackageManagerService.getActivityInfo & 229 & 53.039$\pm$2.223 & 203 & 45.551$\pm$2.104\\\hline
    PackageManagerService.getPackageInfo & 207 & 47.324$\pm$2.339 & 307 & 37.774$\pm$1.456\\\hline
    PackageManagerService.queryIntentServices & 123 & 131.744$\pm$9.220 & 134 & 106.354$\pm$6.069\\\hline
    PackageManagerService.getPackageUid & 93 & 35.767$\pm$2.353 & 201 & 30.005$\pm$0.000\\\hline
  \end{tabular}
  \caption{Ten most frequently invoked hooked functions and their average performance overhead on \OURNAME vs.~stock Android v4.3. The margins of error are given for the 95\% confidence interval.}
  \label{tab:eval:tophooks}
\end{table*}

\section{Module Performance}
\label{sec:appendix:moduleperf}

\begin{figure*}[h]
  \centering
  \includegraphics[width=.6\linewidth]{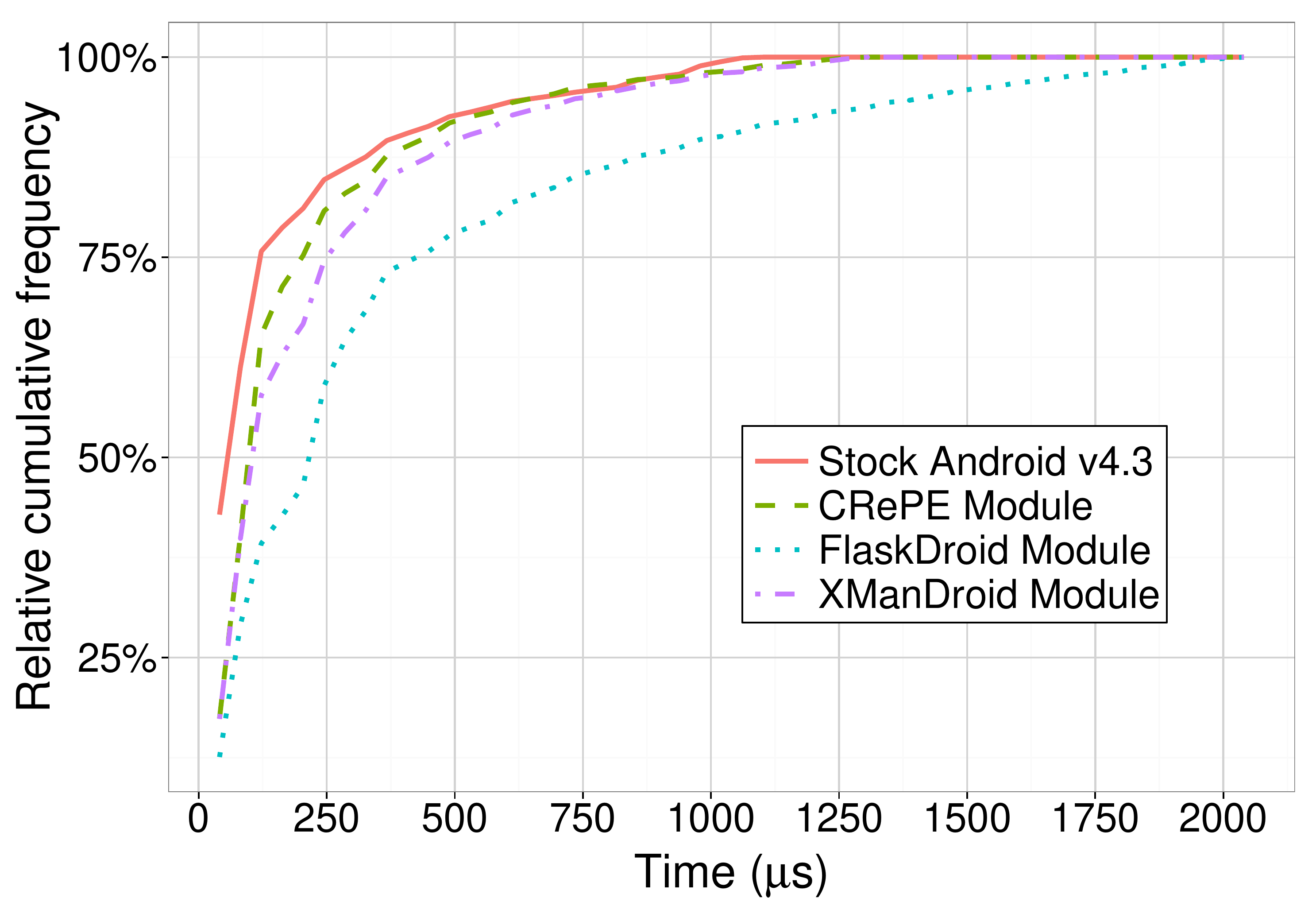}
  \caption{Relative cumulative frequency distribution of example modules' performance overhead vs.~stock Android v4.3.}
  \label{fig:appendix:cfdmodules}
\end{figure*}

Table~\ref{tab:moduleperf} provides an overview of the performance impact of different security models as reported in their respective publications (\textit{native}) and as measured by us for their implementation as module (\textit{\OURSHORT Module}). However, it should be noted, that these are not directly comparable, because all security models have originally been implemented for a different Android OS version and been tested on a different hardware platform. Figure~\ref{fig:appendix:cfdmodules} presents the cumulative frequency distribution for the measured performance overhead of our example modules versus stock Android~v4.3.

\begin{table*}[h]
  \centering
  \begin{tabular}{|l|l|l|l|l|}\hline
    \textbf{Use-case} & \textbf{Implementation} & \textbf{Android version} & \textbf{Test device} & \textbf{Average ($\mu s$)}\\\hline\hline
\multirow{2}{*}{CRePE~\cite{CoNgCr_10:CRePE}}${\star}$ & Native & v2.3 & HTC Magic & $\approx 100$ \\\cline{2-5}
 & \OURSHORT Module$\ ^{\ddagger}$ & v4.3 & Nexus 7 & 168.943$\pm$5.884 \\\hline\hline
\multirow{2}{*}{XManDroid~\cite{BuDaDm_12:TowardsT}} & Native$\ ^{\diamond}$ & v2.2 & Nexus One & $532$ \\\cline{2-5}
& \OURSHORT Module$\ ^{\ddagger}$ & v4.3 & Nexus 7 & 206.062$\pm$5.573\\\hline\hline
\multirow{2}{*}{FlaskDroid~\cite{TUD-CS-2013-0115} (middleware)$^{\dagger}$} & Native & v4.0.3 & Galaxy Nexus & $452$ \\\cline{2-5}
& \OURSHORT Module$\ ^{\ddagger}$ & v4.3 & Nexus 7 & 359.317$\pm$11.015 \\\hline
  \end{tabular}
\captionsetup{
  justification=centering,
}
  \caption{Performance measurements of our example modules.\\
    {\small\textnormal{$^\star$ Two rules loaded.}} {\small\textnormal{$^\diamond$ Weighted average for cached and uncached checks.}} {\small\textnormal {$^\dagger$ With basic policy loaded.}}\\{\small\textnormal {$^{\ddagger}$ Weighted average incl.~IPC roundtrip between hook and module.}} }
  \label{tab:moduleperf}
\end{table*}

\twocolumn

\section{Policy-agnostic Calls from Zygote to Linux Security Module}
\label{sec:appendix:kmacexample}

We briefly explain at the example of Zygote how the generic interface for calls to the kernel module can be used. As described in Section~\ref{sec:architecture}, we added a generic interface implementation, called \texttt{KMAC.java}, to the Android API. \texttt{KMAC.java} implements the interface described in Listing~\ref{listing:kmacapi} in Appendix~\ref{sec:appendix:api}. \texttt{KMAC.java} in turn loads via the Java reflection API the \texttt{LSM.java} classes and via JNI the \texttt{liblsm.so} deployed by modules and uses them to forward calls to the kernel module. \texttt{LSM.java} must hence also implement the interface in Listing~\ref{listing:kmacapi}. \texttt{LSM.java} and \texttt{liblsm.so} are responsible for translating the function arguments to the kernel module specific protocol. For instance, consider Listing~\ref{listing:arch:zygotelsm} that shows how the \texttt{KMAC} interface is used in Zygote to verify that the caller is allowed to specify certain parameters like the UID/GID of a new app process. It first uses the \texttt{getPeerContext} function to retrieve the kernel-level security context of the calling process. This information is stored in a generic Bundle structure. It afterwards uses this information to request policy decisions from the Linux security module such as \texttt{security\_zygote\_applyUidSecurityPolicy} (a Zygote-specific hook). Listing~\ref{listing:arch:selinuxlsm} shows how the SE~Android module (cf.~Section~\ref{sec:usecases}) implements the interface to translate the arguments to SELinux-specific arguments and to call the SELinux kernel module. Here, \texttt{SELinux.java} takes the role of \texttt{LSM.java} and \texttt{SELinux.getPeerContext} and \texttt{SELinux.checkSELinuxAccess} are \textit{native} functions that call via \texttt{libselinux.so} the kernel module. Hence, \texttt{libselinux.so} takes the role of \texttt{liblsm.so}.

\begin{lstlisting}[basicstyle=\footnotesize, caption={Use of generic Kernel module interface in \texttt{ZygoteConnection.java}}, label={listing:arch:zygotelsm},aboveskip=\medskipamount]
private final Bundle peerSecurityContext;
private static  final KMAC mKMAC = new KMAC();
...
ZygoteConnection(LocalSocket socket) throws IOException {
...
  peerSecurityContext = mKMAC.getPeerContext(mSocket.getFileDescriptor());
...
}
...
private static void applyUidSecurityPolicy(Arguments args, Credentials peer, Bundle peerSecurityContext) {
...
  boolean allowed = mKMAC.security_zygote_applyUidSecurityPolicy(peer, peerSecurityContext);
...
}
\end{lstlisting}

\begin{lstlisting}[basicstyle=\footnotesize, caption={Implementation of generic LSM interface for SELinux kernel module}, label={listing:arch:selinuxlsm},aboveskip=\medskipamount]
package android.os;

public class SELinuxAdaptor implements KMACAdaptor {
...
  @Override
  public Bundle getPeerContext(FileDescriptor fd) {
    String ctx = SELinux.getPeerContext(fd);
    Bundle ret = new Bundle();
    ret.putString("selinux.context", ctx);
    return ret;
  }

  @Override
  public boolean security_zygote_applyUidSecurityPolicy(Credentials creds, Bundle peerSecurityContext) {
    String peerCtx = peerSecurityContext.getString("selinux.context");
    return SELinux.checkSELinuxAccess(peerCtx, peerCtx, "zygote", "specifyids");
  }
...
}
\end{lstlisting}

\section{Usage of Instrumentation API}
\label{sec:appendix:irm}

We use the abstract example shown in Listing~\ref{listing:arch:exampleirm} to illustrate the usage of our instrumentation API. Calling \texttt{Instrumentation.redirectMethod()} (line~\ref{listing:callredirect}) diverts the control from method \texttt{foo()} of class \texttt{com.test.A} to the method \texttt{bar()} of class \texttt{com.test.B}. It returns a reference to the original function, which we store in a variable \texttt{A\_foo}. Subsequent calls to \texttt{A.foo()} (line~\ref{listing:callA}) will invoke \texttt{B.bar()} instead. The original method \texttt{foo()} can still be invoked by calling \texttt{Instrumentation.callOriginalMethod(A\_foo)} (line~\ref{listing:callOrigB}).

\begin{lstlisting}[caption={Usage of the IRM instrumentation library}, label={listing:arch:exampleirm},basicstyle=\footnotesize,escapeinside={(*@}{@*)}]
public static void main(String[] args) {
	A.foo(); // calls A.foo()

	MethodHandle A_foo = Instrumentation.redirectMethod( (*@\label{listing:callredirect}@*)
		"com.test.A->foo()",
		"com.test.B->bar()"
	);
  // calls B.bar():
	A.foo(); (*@\label{listing:callA}@*)
  // calls A.foo():
	Instrumentation.callOriginalMethod(A_foo); (*@\label{listing:callOrigB}@*)
}
\end{lstlisting}

\section{Details on FlaskDroid Hook Logic}
\label{sec:flaskdroiddetail}

We illustrate at the example of FlaskDroid~\cite{TUD-CS-2013-0115} how the hook logic of existing solutions can be moved into a module. Listing~\ref{listing:flaskdroidorig} shows one of the original FlaskDroid hooks in the \texttt{getAllProviders} function of the Android location service. The hook calls via the service's context to FlaskDroid's policy server where the access control decision is determined by the \texttt{checkPolicy} function. This function internally determines the subject's and object's security type from their UIDs\footnote{This is the original FlaskDroid behavior that is, as mentioned in Section~\ref{sec:usecase:flaskdroid}, flawed.}. A denial of access results always in a security exception that is thrown back to the caller of the location service API. Listing~\ref{listing:flaskdroidreimplementation} shows the re-implementation of this logic in a module for our \OURNAME by simply overriding the corresponding enforcement function for \texttt{LocationManagerService.getAllProviders} and directly calling the \texttt{checkPolicy} function with all required parameters provided by the hook. It should be noted that in FlaskDroid the security exception in case of denied access is hardcoded in the system, while in an implementation as a module the FlaskDroid authors could alternatively change the enforcement to a less interruptive enforcement by reassigning the \texttt{providerList} parameter to an empty list of Strings, i.e., pretending to the calling app that there is no location provider present in the system. In fact, as a module, such a change in strategy can be more easily rolled out than as a hardcoded implementation within the middleware.

\begin{lstlisting}[language=Java,basicstyle=\footnotesize,caption={Original FlaskDroid hook in \texttt{com.android.server.LocationManagerService}},label={listing:flaskdroidorig},aboveskip=\medskipamount,belowskip=\medskipamount]
public List<String> getAllProviders() {
...
  if(mContext.checkSecurityContext(Binder.getCallingUid(), Process.myUid(), "locationService_c", "getAllProviders") != PackageManager.PERMISSION_GRANTED) {
    throw new SecurityException("Denied by MAC policy");
  }
...
\end{lstlisting}

\begin{lstlisting}[language=Java,basicstyle=\footnotesize,caption={Re-implementation of the hook from Listing~\ref{listing:flaskdroidorig} in a security module},label={listing:flaskdroidreimplementation},aboveskip=\medskipamount,belowskip=\medskipamount]
@Override
public void security_location_getAllProviders(List<String> providerList, int uid, int pid) {
  if(checkPolicy(uid, Binder.getCallingUid(), "locationService_c", "getAllProviders") == PackageManager.PERMISSION_DENIED) {
    throw new SecurityException("Denied by MAC policy");
  }
}
\end{lstlisting}

\section{Further Security Modules}
\label{sec:appendix:furtherusecases}

In this Section we briefly explain further example use-cases from related work that we ported as security modules in case their source code was available to us or that we re-implemented according to their published descriptions.

\subsection{AppOps and IntentFirewall}
\label{sec:usecase:appops}

Google introduced (unofficially) with Android v4.3 the \textit{AppOps} infrastructure for dynamic, more fine-grained Permissions. It introduced hooks in different system services and apps, which query a central AppOpsService whether an application is allowed to perform an operation (e.g., retrieving the location of the device or querying a ContentProvider). The AppOps rules define a mapping from UID/package name to allowed operations. AppOps offers an interface to apps to retrieve the current configuration. Additionally, Google introduced (again unofficially) an \textit{IntentFirewall}, which acts as a reference monitor for certain Intent-based operations like starting an Activity. The IntentFirewall rules describe which caller is allowed to receive which kind of Intent object, using the Intent's attributes such as destination component. The sending or processing of Intents that violate these rules is aborted.

\begin{figure}[h]
  \centering
  \includegraphics[width=.6\linewidth]{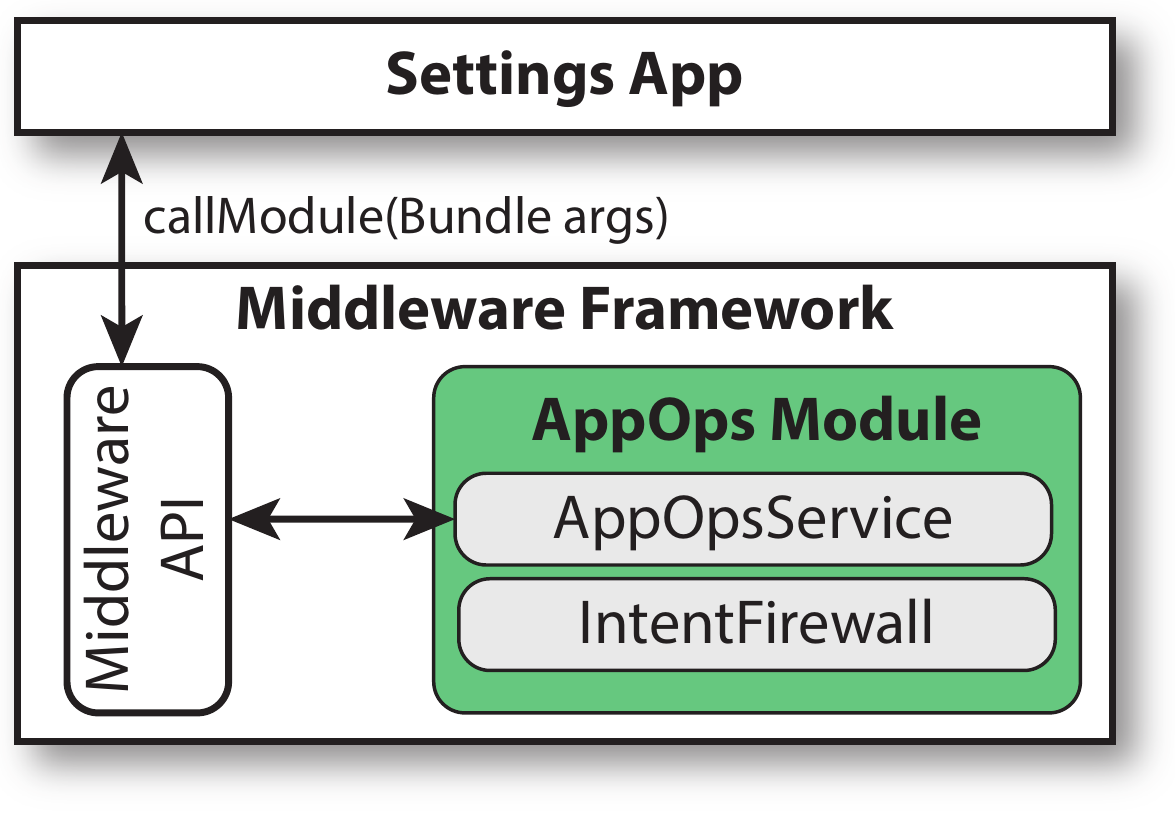}
  \caption{AppOps and IntentFirewall module}
  \label{fig:appopsmodule}
\end{figure}

\paragraph{Implementation as a module:} We ported AppOps and IntentFirewall (from Android v4.3) to a security module for \OURNAME (cf.~Figure~\ref{fig:appopsmodule}) by moving the AppOpsService and the IntentFirewall classes into a module. Our module comprises 2290 lines of code and differs in 33.71\% of all LoC from the native implementation. The bulk of the changes (520 LoC), were required to  move the hook logic of both services from the system apps and services of Android into the module by using our enforcement functions. For the IntentFirewall, this was straightforward and we only had to substitute a direct callback from IntentFirewall to the \textit{ActivityManagerService} by our \OURSHORT callback mechanism. For the AppOpsService, we had to add a mapping from caller PID to package name. By default the hooks of AppOps determine the caller's package name and pass this information to the AppOpsService for policy check. Since this is a policy-specific logic of the hooks, our framework hooks do not (by default) provide the caller's package name and we re-implemented this logic in our module by using our callback interface, which allows us to retrieve the package name for app PIDs. Moreover, we adapted the AppOpsService interface to retrieve/configure the current policies to a Bundle-based communication (e.g., we enabled the \textit{PackageOps} and \textit{OpEntry} to be serialized into a Bundle). AppOps is, furthermore, partially integrated into the \textit{Settings} application to allow users to disable notifications from selected apps. We replaced this policy-specific channel between Settings and AppOps also with our policy-agnostic Bundle-based communication. Modules that support this Settings option, can return a value indicating whether notifications are disabled or not. If the module does not support this feature, Settings app by default allows notifications. However, our AppOps module does currently not support the operation watching feature, which requires the registration of application callback objects with the module.

\subsection{Saint~\cite{OnMcEnMc_09:Saint}}

Saint is an extension for Android OS v1.5 that allows app developers to ship their apps with policy rules that determine how the app can interact with other apps in the system. For instance, the rules can declare that only apps with a specific package name, version, or set of permissions are allowed to call the app or be called by the app. The rules also support defining Intent attributes as rule criteria. The rules are enforced by the system through hooks in different system apps and services, such as the \textit{ActivityManagerService}, that allow monitoring operations for the different app component types. Additionally, Saint provides a front-end app (\textit{FrameworkPolicyManager}), that allows the user to override developer policies.

\paragraph{Implementation as a module:} We \textit{re-implemented} Saint as a security module by developing a module (729 LoC) that supports Saint's policy language as described in \cite{OnMcEnMc_09:Saint}. We use our event functions to extract policy files from newly installed application packages and insert them into a policy database in our module. We use different hooks in the \textit{ActivityManagerService} (e.g., starting an Activity, resolving an Activity, finding active services), Broadcast subsystem, or ContentProvider class to enforce the Saint \textit{runtime} policies. Using the \texttt{scanPackage} hook in the \textit{PackageManagerService} we enforce Saint \textit{install-time} policies to decided whether a new app is installed. Communication between the module the front-end app is again implemented based on Bundles. We successfully verified our Saint module's effectiveness using the policies for the running example described by the Saint author's~\cite{OnMcEnMc_09:Saint} and a set of test apps that implement Saint's example scenario.

\subsection{TrustDroid~\cite{trustdroid}}

TrustDroid extends the Android OS v2.2 architecture with isolation of different domains such as ``work'' and ``private''. Every application is classified during installation into one of the available domains. For classification, TrustDroid uses package-specific attributes such as developer signature, external signature, or package name. Enforcement hooks at middleware and kernel level prevent at runtime any communication between different domains. At kernel level, TrustDroid uses TOMOYO Linux to enforce the policies. Policy rules for newly classified apps are propagated from the middleware to the kernel.

\paragraph{Implementation as a module:} To \textit{re-implement} TrustDroid as a security module (862 lines of code), we deployed a TOMOYO-enabled Linux kernel on the device (i.e., our \KMODULE) and developed a \MMODULE that deploys the required \texttt{LSM.java} and \texttt{libccs.so} to communicate with the kernel module. Additionally, we used our \texttt{scanPackage} hook in the \textit{PackageManagerService} to classify newly installed applications and keep a mapping from UID to domain.\footnote{TrustDroid does not allow apps in a shared sandbox to be classified differently.} Because TrustDroid's policy is static and very simple, its architecture does not distinguish between policy enforcement and policy decision point, but instead every hook retrieves the domain of the current subject and object and denies access if their domains differ. We re-implemented this logic using the enforcement functions of our module, which was a straightforward implementation. For ContentProviders (e.g., Contacts) TrustDroid classifies the database entries and returns on access only the entries that have the same domain as the caller. Since we prohibit by design such as policy-specific intrusions into the default ContentProviders, we use our pre-query hooks to modify selection arguments to retrieve only contacts that are allowed for the current caller (e.g., where the contact's group indicates a private contact). Using two example applications that are classified differently, we verified the effectiveness of our TrustDroid module.

\subsection{Data shadowing~\cite{HoHaJuScWe_11:RetrofittingAndroid,tissa11}}

Both \textit{AppFence}~\cite{HoHaJuScWe_11:RetrofittingAndroid} as well as \textit{TISSA}~\cite{tissa11} provide a data shadowing feature. Data shadowing means, that an application that wants to retrieve sensitive information (e.g., contacts information, IMEI number, or location data) only get empty, fake, or filtered data.

\paragraph{Implementation as a module:} We \textit{re-implemented} the data shadowing features of AppFence and TISSA as a module by using our \textit{edit automaton} hooks in the \texttt{ContentProvider.Transport} class, the ContactsProvider-specific hooks, Telephony service and Location service. For ContentProvider and ContactsProvider, in particular our pre-query and post-query hooks allowed us a fine-grained filtering or replacing (faking) of the returned data as well as returning an empty data set. However, the current coverage of our enforcement hooks does not include some of the data shadowing points of AppFence, such as microphone, logs, or camera, and we plan on adding them in the future.

\subsection{Kirin~\cite{EnOnMc_09:Kirin}}

Kirin extends Android's application installation process with policy-based checks and denies installation of a new app when it violates the policy. Based on its time of publication, we presume that it was developed for Android OS v1.5.\footnote{\url{http://en.wikipedia.org/wiki/Android_version_history}} The actual policy check was performed in a dedicated Android application developed for Kirin, which interacted with the installation process. These policies are based on the set of permissions requested by an app and the interfaces (e.g., Broadcast receivers) it wants to register in the system. The installation of apps that are rejected by the policy is denied.

\paragraph{Implementation as a module:} To \textit{re-implement} Kirin's security service as a security module, we developed a module that supports Kirin's security language. Using our \texttt{scanPackage} hook in the \textit{PackageManagerService}, we check new applications against the policy and abort their installation in case the policy rejects the application. Our Kirin module comprises 246 lines of code.

\end{document}